\documentclass[twocolumn,superscriptaddress,prb]{revtex4-1}

\usepackage[T1]{fontenc}
\usepackage[utf8]{inputenc}
\usepackage{lmodern}
\usepackage{microtype}
\usepackage{amsmath,amssymb,bm,mathtools}
\usepackage{graphicx}
\usepackage{booktabs,array}
\usepackage[colorlinks=true,citecolor=blue,urlcolor=blue,linkcolor=blue]{hyperref}
\usepackage[noabbrev]{cleveref}
\setcitestyle{numbers,square}

\newcommand{\ii}{\mathrm{i}}
\newcommand{\e}{\mathrm{e}}
\newcommand{\dd}{\mathrm{d}}
\newcommand{\bbZ}{\mathbb Z}
\newcommand{\bbE}{\mathbb E}
\newcommand{\cF}{F}
\newcommand{\cR}{\mathcal R}
\newcommand{\Tr}{\operatorname{Tr}}
\newcommand{\ket}[1]{\lvert #1\rangle}
\newcommand{\bra}[1]{\langle #1\rvert}
\newcommand{\doilink}[1]{\href{https://doi.org/#1}{doi:#1}}
\newcommand{\arxiv}[1]{\href{https://arxiv.org/abs/#1}{arXiv:#1}}

\begin{document}

\title{A Single Conformal Coordinate Decodes Outcome-Resolved Measurement-Induced Entanglement}

\author{Reyhaneh Khasseh}
\affiliation{Theoretical Physics III, Center for Electronic Correlations and Magnetism, Institute of Physics, University of Augsburg, D-86135 Augsburg, Germany}
\author{M.~A.~Rajabpour}
\affiliation{Instituto de F\'isica, Universidade Federal Fluminense,
Av.~Gal.~Milton Tavares de Souza s/n, Gragoat\'a, 24210-346, Niter\'oi, RJ, Brazil}
\affiliation{Theoretical Physics III, Center for Electronic Correlations and Magnetism, Institute of Physics, University of Augsburg, D-86135 Augsburg, Germany}
\date{\today}

\begin{abstract}
A local measurement of a critical many-body state produces exponentially many
microscopic outcomes, each defining a different conditioned quantum state.
We ask how much information from one such outcome is actually needed to predict
the entanglement that remains between two unmeasured regions.  We show that,
in critical quantum chains, the universal outcome-resolved R\'enyi entropy is
asymptotically determined by a single compact conformal coordinate extracted
directly from the measurement record.  Compact-boson theory explains this
compression: the universal entropy is insensitive to all spatially varying
components of the boundary profile, retaining only a single relative compact
zero mode.  At the free-fermion XX point, exact lattice identities expose the
microscopic structure underlying the decoder, while large-scale
outcome-by-outcome calculations confirm its scaling prediction across
different subsystem geometries and R\'enyi indices.  Numerical calculations
in interacting XXZ chains further show that the same microscopic coordinate
continues to decode the conditional entropy without interaction-dependent
retuning.  Thus an exponentially large space of measurement outcomes is
reduced, for this universal observable, to a single conformal degree of
freedom, revealing an emergent Luttinger-liquid mechanism for
outcome-resolved postmeasurement entanglement.
\end{abstract}

\maketitle

\section{Introduction}
A projective measurement produces not only a new quantum state but also a
classical record with microscopic spatial structure.  Site-resolved snapshots
are now routinely accessible in quantum-gas microscopes
\cite{Bakr2009,Sherson2010,Cheuk2015,Parsons2015}, while monitored quantum
processors have made measurement-induced entanglement, and
critical phenomena experimentally accessible
\cite{Noel2022,GoogleAI2023,Koh2023}.  On the theoretical side, measurements
are known to generate criticality, purification, and nontrivial conformal and
information-theoretic structure in both monitored circuits and critical
many-body states
\cite{LiChenFisher2018,Skinner2019,Li2019,GullansHusePRX2020,
LiChenLudwigFisher2021,Zabalo2022,Agrawal2022,Fava2023,
LeeJianXu2023,Fisher2023,
Minoguchi2022,Buchhold2021,Garratt2023,Murciano2023,Yang2023,
SunYaoJian2023,Ashida2024,Paviglianiti2023}.
Recent work has further emphasized boundary phenomena, intrinsic measurement
randomness, and universal information contained in individual snapshots
\cite{LiuMurcianoMrossAlicea2025,TangWen2025,PatilLudwig2024,
PatilLudwig2025,Baweja2024,SarmaSnapshots2026,Naus2025}.
This motivates a distinctly outcome-resolved question: how much information
from a single microscopic measurement record must be retained to predict a
universal property of the corresponding conditioned state?

Measurement-induced entanglement (MIE) provides a natural setting for this
question.  A region $C$ is measured while two separated regions $A$ and $B$
remain unmeasured, linking localizable and measurement-prepared entanglement
\cite{Verstraete2004,Popp2005,Lu2022,Tantivasadakarn2024} with information
diagnostics, postselection-free probes, and complexity
\cite{Lin2023,Cheng2024,GarrattAltman2024,McGinleyLearning2024,McGinley2025}.
Boundary-CFT calculations established the slit geometry associated with
selected measurement outcomes
\cite{Rajabpour2015,Rajabpour2016,NajafiRajabpour2016,Hoshino2025,
Stephan2014,RajabpourFormation2015}.  Tomonaga--Luttinger liquids admit a
compact-boson description
\cite{Tomonaga1950,Luttinger1963,Haldane1981a,Haldane1981b,Giamarchi2003},
and recent work has determined the Born-averaged MIE and its statistics
\cite{KhannaVasseur2026,KhannaVasseurStats2026}.  These ensemble results,
however, do not determine the joint relation between a particular measurement
record and the entanglement of the state conditioned on that same record.
Recent free-fermion work further showed that, in the CFT regime, the
conditional entropy can be parametrized by a single induced charge of the
postmeasurement state, even though the full entanglement Hamiltonian retains
additional outcome dependence \cite{EislerTonni2026}.  Here we instead ask
whether the entropy-relevant variable can be read directly from the microscopic
measurement record itself, without first reconstructing the conditioned state.

For the critical XX chain, Ref.~\cite{RajabpourSnapshots2026} associated each
partial occupation snapshot $m$ with a compact conformal coordinate
$\delta_L(m)\in[0,2\pi)$ and determined its Born distribution.  For a given
measurement record $m$, let $S_{\alpha,L}(m)$ denote the R\'enyi entanglement
entropy between the two unmeasured regions in the corresponding conditioned
state.  Here we establish the stronger outcome-by-outcome relation
\begin{equation}
 S_{\alpha,L}(m)\simeq F_\alpha^{(g)}\!\left(h,\delta_L(m)\right),
 \label{eq:main-decoder}
\end{equation}
where $F_\alpha^{(g)}$ is the universal compact-boson prediction.  Thus, for
Born-typical records in the scaling limit, the coordinate extracted from a
particular measurement record predicts the R\'enyi entropy of the state
conditioned on that same record.  The four-interval geometry enters through
the cylinder height $h$, while the low-energy interaction dependence is
encoded in the compact-boson coupling $g$.

Two features make Eq.~\eqref{eq:main-decoder} nontrivial.  First, a typical
measurement record remains rough down to the lattice scale, so replacing it
from the outset by a uniform conformal boundary condition would discard
precisely the microscopic information whose relevance is being tested.
Second, at the XX point the microscopic problem admits exact free-fermion
identities and a two-cut Coulomb-gas representation, providing an analytical
route to the compact-boson scaling description, whereas the underlying
continuum mechanism is not restricted to the free-fermion coupling. This separation leads to a stringent
universality question: does the \emph{same microscopic coordinate} continue to
decode the entropy once the free-fermion structure is removed?  We address
this in the periodic spin-$1/2$ XXZ chain,
\begin{equation}
H(\Delta)=\sum_j(S_j^xS_{j+1}^x+S_j^yS_{j+1}^y+
\Delta S_j^zS_{j+1}^z), 
\end{equation}
at zero magnetization, with the XX point at
$\Delta=0$ and critical regime $-1<\Delta\le1$.  For this family,
$g(\Delta)=\arccos(-\Delta)/\pi$
\cite{Haldane1981a,Haldane1981b,Giamarchi2003}, with $g(0)=1/2$.
Exact diagonalization and matrix-product-state calculations then test the
outcome-resolved relation in the interacting regime without redefining
$\delta_L$ or fitting the interaction dependence.

\begin{figure}[t]
\centering
\includegraphics[width=0.8\columnwidth]{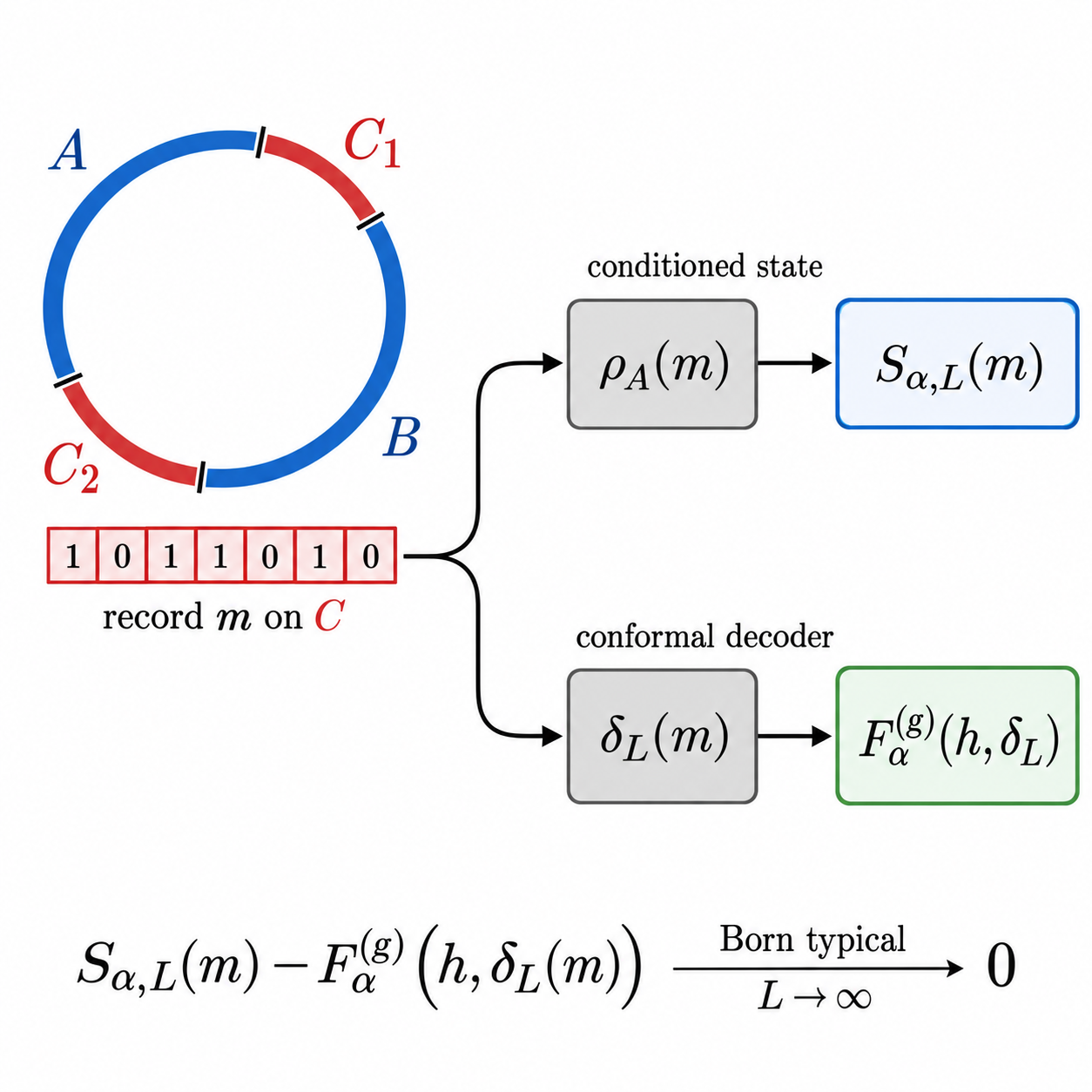}
\caption{\textbf{Outcome-resolved conformal decoder.}
A single measurement record $m$ defines two independent routes.
Conditioning the many-body state gives the outcome-resolved R\'enyi entropy
$S_{\alpha,L}(m)$, while the measured bits independently determine the compact
conformal coordinate $\delta_L(m)$ and hence the continuum prediction
$F_\alpha^{(g)}(h,\delta_L)$.  The central result is that these two quantities
agree for Born-typical outcomes in the scaling limit.}
\label{fig:decoder}
\end{figure}

\section{Outcome-resolved conformal decoder}
\subsection{Outcome-resolved conformal coordinate}
Four consecutive intervals $A,C_1,B,C_2$ partition the periodic chain, with
only $C=C_1\cup C_2$ measured in the $S^z$ (occupation) basis,
Fig.~\ref{fig:decoder}.  Their ordered endpoints define the standard slit
geometry and an equivalent cylinder of circumference $2\pi$ and height $h$
\cite{Rajabpour2015,Rajabpour2016,Stephan2014}.  Let $W=u+iv$ denote the
corresponding conformal map and define the harmonic side coordinate
$f=2\,\mathrm{Im}\,W$.  It is constant on the two unmeasured intervals and
winds with opposite orientations along $C_1$ and $C_2$.  Sampling $f$ at the
lattice midpoints gives the fixed geometric weights $f_j$.  The snapshot
coordinate introduced in Ref.~\cite{RajabpourSnapshots2026} is
\begin{equation}
 \delta_L(m)=\left[\sum_{j\in C}f_j\left(m_j-\frac12\right)-\pi|B|\right]_{2\pi}.
 \label{eq:delta-pr}
\end{equation}
This is neither a particle count nor a simple imbalance: the conformal weights
are fixed entirely by the four endpoints and select a specific long-wavelength
component of the measurement record.  In the continuum, $\delta_L$ becomes the
relative angular average of the two boundary profiles and, being defined
modulo $2\pi$, is determined entirely by the measured bits.

This construction cleanly separates geometry from dynamics.  Once the four
endpoints are fixed, the weights $f_j$ are determined entirely by the geometry. Changing the interaction therefore
leaves the microscopic coordinate $\delta_L(m)$ unchanged, while the continuum
prediction changes only through the known coupling $g$.  Conversely, changing
the interval geometry modifies the harmonic weights and the cylinder height
$h$ in a prescribed way.  The decoder is thus not a generic scalar summary of
the measurement record, but a geometrically selected coordinate whose
entanglement dependence is fixed by the compact-boson theory.  The microscopic
conventions, conformal geometry, and exact conditional correlation matrix are
collected in Appendix~\ref{sec:model}.

\subsection{Continuum mechanism}
The continuum derivation does not require either boundary profile to be smooth
or uniform.  Consider a compact boson $\phi\equiv\phi+2\pi$ with action
$S=(g/4\pi)\int |\nabla\phi|^2$ on the cylinder, with arbitrary periodic
Dirichlet profiles $\varphi_1(v)$ and $\varphi_2(v)$
\cite{Cardy1984,Cardy1989}.  Decompose each profile into its angular average
and nonzero Fourier modes.  On the connected $n$-sheet cover
\cite{CalabreseCardy2004}, the cylinder height becomes $h/n$ while a boundary
mode $k$ is mapped to mode $nk$.  The classical action of every nonzero mode
is therefore multiplied by exactly $n$, matching the contribution from the
$n$ one-sheet normalizations.  All nonzero Fourier components consequently
cancel from the normalized replica ratio before any Born average is taken.

The compact zero-mode sector is different.  The connected surface contains a
single winding sum, whereas the denominator contains $n$ independent winding
sums, as in the compact-boson MIE construction of
Refs.~\cite{KhannaVasseur2026,KhannaVasseurStats2026}.  Writing
$\delta=[\varphi_{2,0}-\varphi_{1,0}]_{2\pi}$ gives
\begin{align}
 \mathcal R_n^{(g)}(h,\delta)
 &=\frac{\eta(q_1)^n}{\eta(q_n)}
   \frac{\Theta_n^{(g)}(h,\delta)}
        {\Theta_1^{(g)}(h,\delta)^n},
 \label{eq:R-pr}\\
 \Theta_n^{(g)}(h,\delta)
 &=\sum_{w\in\mathbb Z}
 \exp\!\left[-\frac{ng}{2h}(\delta+2\pi w)^2\right].
 \nonumber
\end{align}
Here $q_n=\e^{-2\pi^2n/h}$.  The corresponding universal R\'enyi function is
\begin{equation}
F_n^{(g)}(h,\delta)
=
\frac{1}{1-n}\log \mathcal R_n^{(g)}(h,\delta).
\label{eq:F-pr}
\end{equation}
Thus an arbitrarily nonuniform boundary profile affects the universal
outcome-resolved entropy only through the relative compact zero mode
$\delta$.  This cancellation holds for arbitrary compact-boson coupling and
is profile-by-profile, rather than emerging only after averaging over
measurement outcomes.

\begin{figure*}[t]
	\centering
	\includegraphics[width=0.85\textwidth]{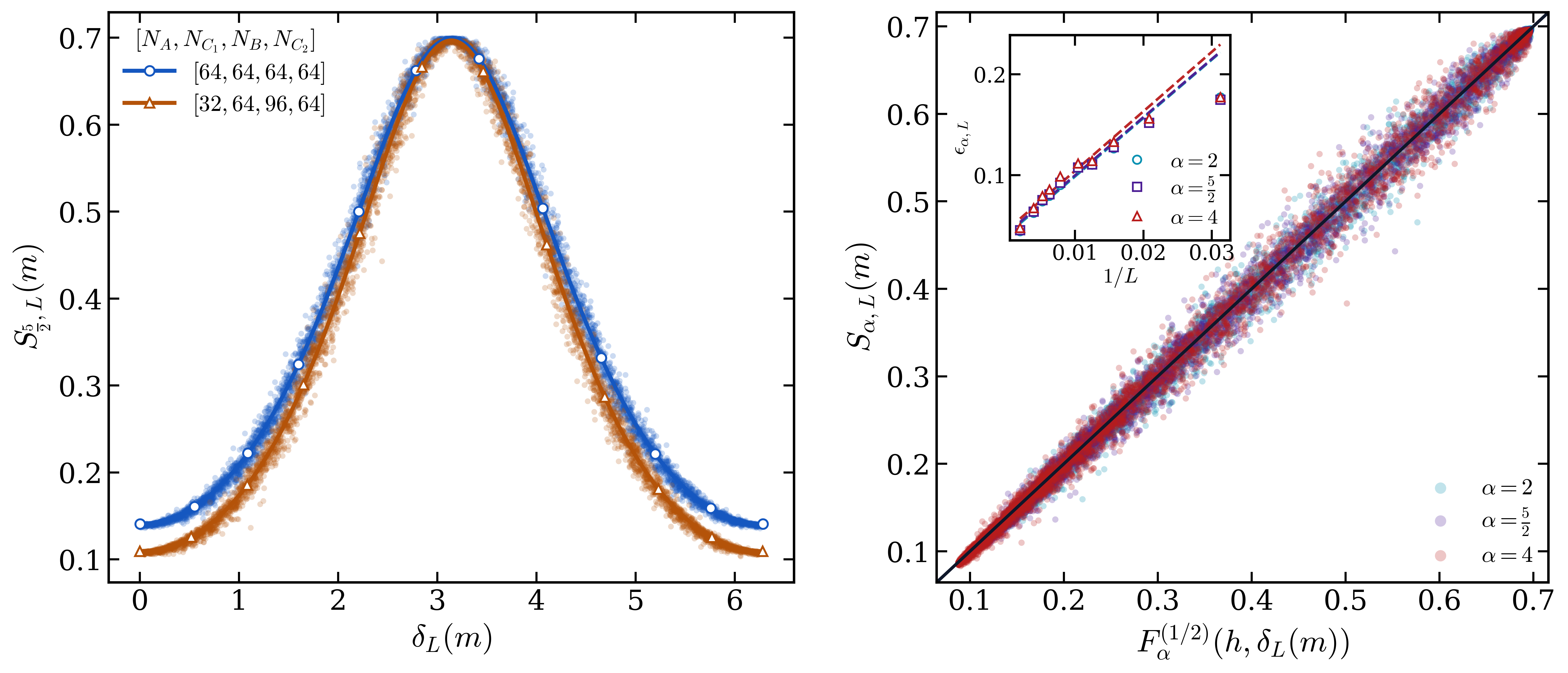}
	\caption{\textbf{Exact XX benchmark of the outcome-resolved decoder.}
		(a) Outcome-resolved entropies at $L=256$ and $\alpha=5/2$ for two distinct
		interval geometries.  Points are exact lattice results and solid curves are the
		parameter-free compact-boson predictions.
		(b) Exact entropy versus continuum prediction for the strongly asymmetric
		geometry at $\alpha=2,5/2,4$; the diagonal is equality, not a fit.  The inset
		shows the finite-size convergence of $\epsilon_{\alpha,L}$ for the same
		R\'enyi indices, with dashed lines as large-$L$ fit guides.
		All numerical data use $5000$ exact Born samples per geometry generated by
		projection-DPP sampling.}
	\label{fig:xx}
\end{figure*}

This mechanism is distinct from boundary locking in R\'enyi-weighted formation
probabilities, where the measurement configurations themselves are reweighted
and the effective boundary condition may change
\cite{StephanMisguichPasquier2011,RajabpourShannonBoundary2026}.
Here the replica construction acts on the conditional density matrix while
the measurement record is held fixed.  For comparison with noninteger
R\'enyi entropies we use the continuation $n\to\alpha$ of
Eq.~\eqref{eq:R-pr}; the lattice entropies are always evaluated directly for each $\alpha$.
The fixed-profile compact-boson construction and the explicit mode-by-mode
Dirichlet calculation are given in Appendices~\ref{sec:cft} and
\ref{app:boson}, respectively.

\subsection{Microscopic XX matching}
The XX chain provides a microscopic explanation for why
Eq.~\eqref{eq:delta-pr}, rather than a generic scalar statistic of the
measurement record, is the relevant variable.  After the Jordan--Wigner
transformation, the half-filled ground state is a free-fermion Slater
determinant, equivalently a projection determinantal process
\cite{JordanWigner1928,LiebSchultzMattis1961,Katsura1962}.  Conditioning on
measured occupations preserves Gaussianity, so the exact finite-size
entanglement follows from the conditional correlation matrix
\cite{Peschel2003}.  For integer $n$, the same R\'enyi moment admits an exact
full-counting-statistics representation \cite{Klich2003}.  At this stage the
full microscopic record is retained exactly: no averaging over outcomes and
no replacement of the binary snapshot by a smooth boundary profile is
involved.

To make this structure explicit, let $\Pi_m$ project onto the measured
occupation record $m$, and let
$N_A=\sum_{j\in A}c_j^\dagger c_j$ denote the fermion number in $A$.  We
introduce the exact lattice counting function
\begin{equation}
 D_{m,L}(\lambda)
 =
 \bra{\Psi_0}\Pi_m\lambda^{N_A}\ket{\Psi_0}.
\end{equation}
It generates the joint weights of the record $m$ and the particle number in
$A$; in particular,
$D_{m,L}(1)=P_{\rm Born}(m)$, so that
$D_{m,L}(\lambda)/D_{m,L}(1)$ is the counting function in the state
conditioned on that record.  Defining
$\lambda_r=\exp[i\pi(2r+1-n)/n]$, one obtains the exact identity
\begin{equation}
 \Tr\rho_A(m)^n
 =
 \prod_{r=0}^{n-1}
 \frac{D_{m,L}(\lambda_r)}{D_{m,L}(1)}.
 \label{eq:lattice-factor-prl}
\end{equation}
Thus the numerator and denominator contain precisely the same microscopic
measurement record.  The subsequent reduction to a single infrared variable
therefore cannot arise from averaging over outcomes.  The complete exact
replica representation and the conditional-Slater-determinant identities are
given in Appendices~\ref{sec:lattice} and \ref{app:gaussian}, respectively.

In position space, $D_{m,L}$ becomes a fixed-filling sum of
Vandermonde-squared weights on the two unmeasured cuts.  Measured occupied
sites act as pinned logarithmic charges, measured empty sites are excluded,
and the remaining particles occupy $A\cup B$.  The collective infrared
degree of freedom is then the relative filling of the two unmeasured cuts.
In the Coulomb-gas description, the coarse-grained height and logarithmic
potential are harmonic conjugates; pairing this height with the conformal side
coordinate identifies the displacement of the relative-filling mode with the
snapshot coordinate $\delta_L(m)$; the corresponding harmonic-coordinate
derivation is given in Appendix~\ref{app:height}.  At fixed long-wavelength cutoff, the
two-cut filling sectors have the asymptotic weights
\begin{equation}
 \mathcal W_L(s\mid m)\propto
 \exp\!\left[-\frac{\pi^2}{h}
 \left(s+\frac{\delta_L(m)}{2\pi}\right)^2\right],
 \qquad s\in\mathbb Z,
 \label{eq:filling-prl}
\end{equation}
which directly reconstruct the compact theta-function sector of
Eq.~\eqref{eq:R-pr}.  This is the genus-one fixed-filling structure of the
two-cut problem
\cite{BorotGuionnet2024,BorodinGorinGuionnet2017,
BaikKriecherbauerMcLaughlinMiller2003}.

Extending these regulated asymptotics to a binary lattice snapshot requires
that lattice-scale boundary roughness not contribute at $O(1)$ to the
normalized entropy.  The fixed-cutoff matching and the precise ultraviolet
condition, including the response of the omitted modes, are given in
Appendices~\ref{sec:matching} and \ref{app:matching}.  The numerical results below probe the
resulting parameter-free scaling prediction.

Equation~\eqref{eq:main-decoder} is therefore a scaling prediction, not an
unconditional finite-lattice identity.  The conditional Gaussian formulas,
the replica factorization, the pinned two-cut representation, and the
fixed-profile compact-boson cancellation are exact in their respective
descriptions.  Passing from the rough binary snapshot to the continuum
decoder additionally uses the fixed-cutoff multi-cut asymptotics and the
irrelevance of lattice-scale boundary roughness to the normalized entropy.

\section{Numerical tests}
\subsection{Exact XX benchmark}
At the XX point, Born records can be sampled exactly by sequential conditioning
of the Fermi projector.  For each record, $S_{\alpha,L}(m)$ is obtained from
the conditional single-particle spectrum, while $\delta_L(m)$ is computed
independently from the measured bits using Eq.~\eqref{eq:delta-pr}.  No
offset, amplitude, stiffness, or curve shape is fitted.  Figure~\ref{fig:xx}
tests the resulting prediction across geometry, R\'enyi index, and system size.
The exact DPP sampling procedure and additional finite-size diagnostics are
given in Appendix~\ref{sec:xx-numerics}.

To quantify the outcome-by-outcome agreement, we use
\begin{equation}
\begin{aligned}
E_{\alpha,L}
&=
\Big\langle
\big[S_{\alpha,L}(m)
-
F_\alpha^{(g)}(h,\delta_L(m))\big]^2
\Big\rangle_{\rm Born},
\\
\epsilon_{\alpha,L}
&=
\sqrt{\frac{E_{\alpha,L}}
{\operatorname{Var}_{\rm Born}(S_{\alpha,L})}},
\qquad
Q_{\alpha,L}=1-\epsilon_{\alpha,L}^{\,2}.
\end{aligned}
\label{eq:decoder-diagnostics}
\end{equation}
Thus $\epsilon_{\alpha,L}\to0$, or equivalently $Q_{\alpha,L}\to1$,
signals asymptotically exact outcome-resolved decoding.

\begin{figure*}[t]
	\centering
	\includegraphics[width=\textwidth]{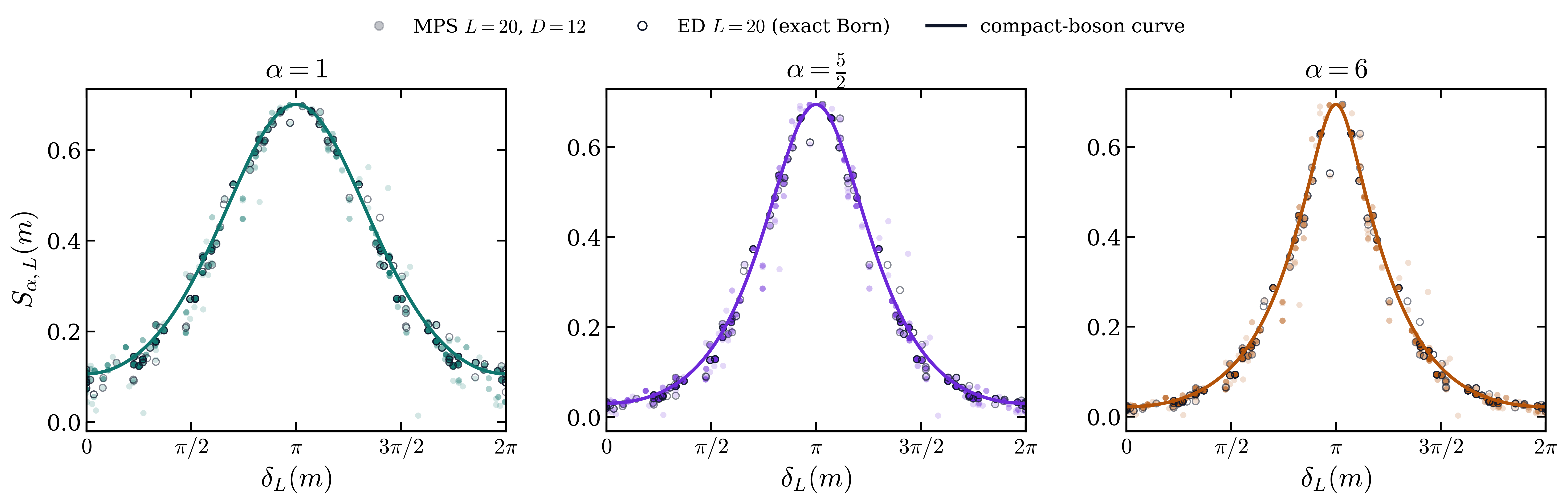}
	\caption{ED--MPS comparison for the interacting XXZ chain at
		$L=20$ and $\Delta=1/2$, for $\alpha=1$, $5/2$, and $6$ (left to right).
		Open circles show exact ED results, while colored points show MPS results
		with bond dimension $D=12$.  Solid curves are the parameter-free
		compact-boson predictions $\cF_\alpha^{(g)}(h,\delta_L)$ with
		$g=2/3$.
	}
	\label{fig:xxz-ed-mps-three-alpha}
\end{figure*}

Figure~\ref{fig:xx}(a) tests the decoder at $L=256$ and
$\alpha=5/2$ for two distinct macroscopic partitions.  Changing the subsystem
geometry changes both the harmonic coordinate $\delta_L$ and the cylinder
height $h$, producing different continuum curves.  Nevertheless, the
individual lattice outcomes follow the corresponding parameter-free
predictions $F_{5/2}^{(1/2)}(h,\delta_L)$ in both geometries.  The agreement is
therefore not tied to the equal-quarter configuration.

Figure~\ref{fig:xx}(b) keeps the strongly asymmetric geometry fixed and
compares the exact entropy directly with
$F_\alpha^{(1/2)}(h,\delta_L)$ for $\alpha=2,5/2,4$.  The outcomes lie close to
the equality line without regression or record-dependent fitting.  The
noninteger case $\alpha=5/2$ provides an independent test of the continuation
away from integer replica index, because the lattice entropy is evaluated
directly from the conditional spectrum.  The inset shows the corresponding
finite-size decrease of $\epsilon_{\alpha,L}$.

\subsection{Interacting XXZ}
The free-fermion benchmark raises a sharper question: does the same microscopic
decoder survive once the determinantal structure is removed?  We test this in
the critical XXZ chain using periodic matrix-product-state calculations.  The
coordinate $\delta_L(m)$ is kept exactly unchanged from the XX construction,
while the continuum prediction uses the known interaction-dependent
Luttinger coupling, with no fitted stiffness or interaction-dependent
redefinition.

Figure~\ref{fig:xxz-ed-mps-three-alpha} provides a direct same-size
validation of the MPS calculation against ED at $L=20$ and
$\Delta=1/2$.  For $\alpha=1$, $5/2$, and $6$, the MPS results closely
reproduce the exact ED outcome-resolved entropies over the full range of
the compact coordinate $\delta_L(m)$.  Both datasets also follow the same
parameter-free compact-boson prediction.  The agreement across the von
Neumann, noninteger, and high-R\'enyi cases shows that the MPS calculation
accurately captures the conditional entanglement structure relevant to
the interacting decoder already at this finite system size.

\begin{figure}[t]
\centering
\includegraphics[width=0.5\textwidth]{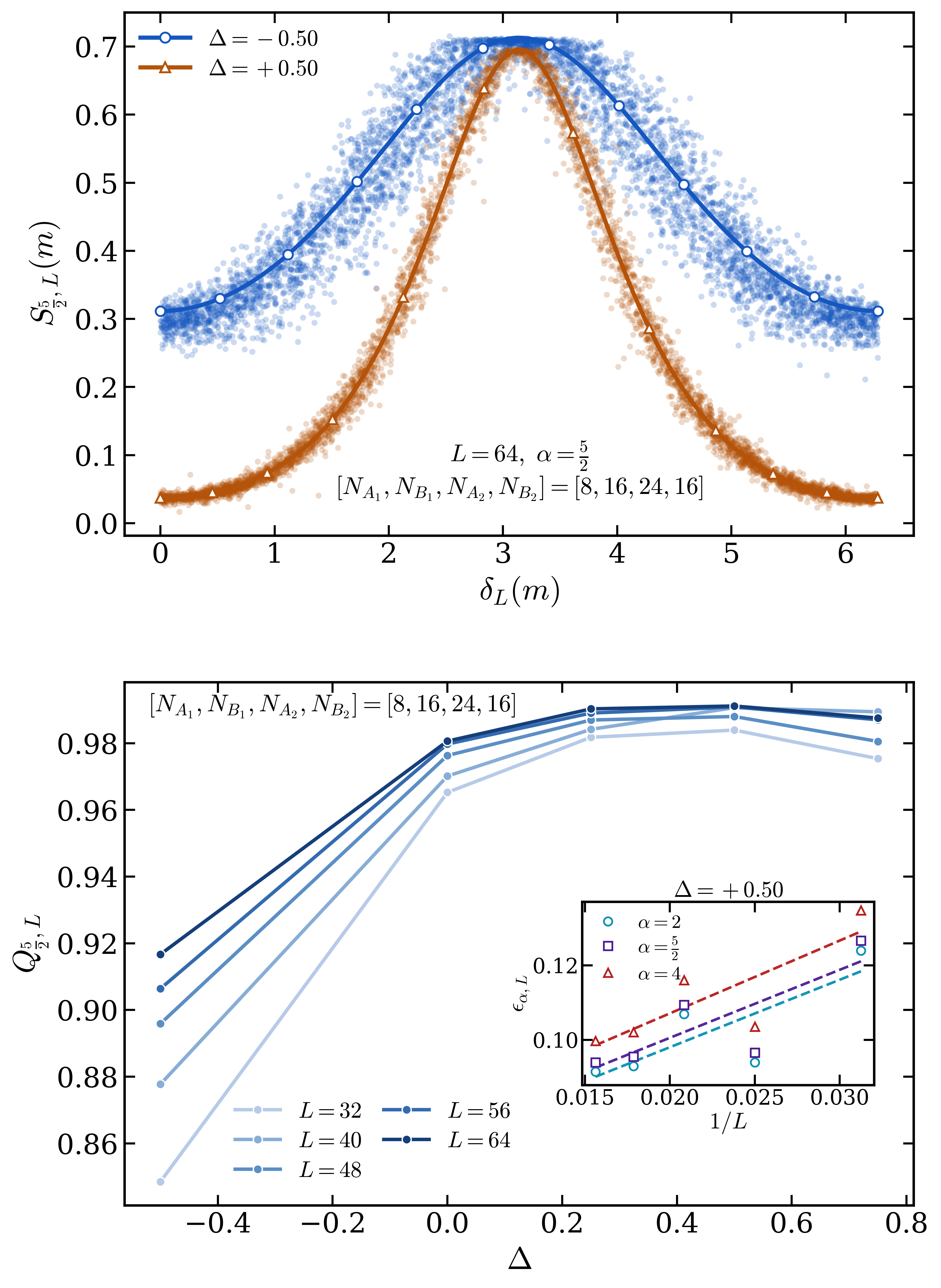}
\caption{\textbf{Outcome-resolved decoder in the critical XXZ chain.}
(a) At $L=64$ and $\alpha=5/2$, results are shown for
$\Delta=-0.50$ and $+0.50$.  Solid curves are the corresponding
parameter-free compact-boson predictions obtained from the same
$\delta_L(m)$.
(b) $Q_{\alpha,L}$ versus anisotropy for several system sizes at
$\alpha=5/2$.  The inset shows $\epsilon_{\alpha,L}$ versus $1/L$ at
$\Delta=+0.50$ for $\alpha=2,5/2,4$; dashed lines are finite-size fit guides. All data are obtained from periodic-MPS Born sampling with bond dimension
$D=32$.}
\label{fig:xxz}
\end{figure}

Figure~\ref{fig:xxz}(a) also gives a direct outcome-by-outcome test away from the
free-fermion point.  At fixed $L=64$, fixed geometry, and $\alpha=5/2$,
changing $\Delta$ deforms the continuum curve through $g(\Delta)$, while the
same lattice coordinate $\delta_L(m)$ continues to organize the conditional
entropies on both sides of the XX point.  Figure~\ref{fig:xxz}(b) extends this
test across the critical XXZ regime: $Q_{\alpha,L}$ remains close to unity over
the tested anisotropies, while the inset shows the finite-size decrease of
$\epsilon_{\alpha,L}$ at fixed $\Delta=1/2$ for several R\'enyi indices.
Together, the two panels test interaction, system-size, and R\'enyi-index
dependence without retuning the microscopic decoder.

Because a critical MPS with finite bond dimension $D$ has a finite
correlation length $\xi(D)$, these finite-size trends are interpreted only
within the regime where $L/\xi(D)$ remains controlled
\cite{Tagliacozzo2008,Pollmann2009}.  The corresponding numerical protocols, and finite-entanglement diagnostics are given in
Appendix~\ref{app:xxz-numerics}.  Unlike the XX derivation, the interacting
calculation does not rely on Gaussian conditioning; what remains common is
only the measured record, the geometrically fixed coordinate $\delta_L(m)$,
and the compact-boson prediction evaluated at the known $g(\Delta)$.
An analytical microscopic matching for XXZ remains open.

\section{Relation to ensemble MIE and consistency checks}
At the XX point, the outcome-resolved relation contains the known ensemble
theory as a direct consequence.  The conformal coordinate has the heat-kernel
distribution $P_h(\delta)$ derived in
Ref.~\cite{RajabpourSnapshots2026}.  Combining this distribution with the
record-level prediction gives
\begin{equation}
 \overline{S}_n(h)
 =
 \int_0^{2\pi}
 P_h(\delta)\,
 F_n^{(1/2)}(h,\delta)\,d\delta,
 \label{eq:ensemble-mie}
\end{equation}
which reproduces the compact-boson Born-averaged MIE of
Ref.~\cite{KhannaVasseur2026}.  Likewise, pushing the same distribution
through $F_n^{(1/2)}$ gives the limiting entropy statistics discussed in
Ref.~\cite{KhannaVasseurStats2026}.  These ensemble results therefore follow
from the outcome-resolved decoder, but not conversely: the same marginal
entropy statistics could in principle coexist with a finite spread of
entropies among outcomes having the same $\delta$.  The collapse in
Figs.~\ref{fig:xx} and \ref{fig:xxz} directly tests this stronger
record-by-record relation.  The full distributional consequences are collected
in Appendix~\ref{sec:consequences}.

The compact coordinate can also be checked independently of the entropy.
Its Born distribution obeys the compact-boson Fourier-moment law; at the XX
point this reduces to the result of
Ref.~\cite{RajabpourSnapshots2026}, while in the interacting XXZ chain the
same test probes whether the unchanged microscopic coordinate acquires the
expected interaction-dependent compact-boson distribution.  The corresponding
moment test and numerical results are given in
Appendix~\ref{app:xxz-numerics}.

Finally, the same zero-mode theory controls two geometric limits.  For a
short cylinder the nearest winding sector dominates, so the entropy remains
sensitive to the compact coordinate, whereas for a long cylinder the
$\delta$ dependence is exponentially suppressed.  In the latter limit,
\begin{equation}
 F_n^{(g)}(h,\delta)
 =
 \frac{h}{12}\left(1+\frac{1}{n}\right)
 -\frac{1}{2}\log(2g)
 +O(e^{-c_{n,g}h}),
 \label{eq:long-cylinder}
\end{equation}
independently of $\delta$ at leading order.  Thus varying the conformal
geometry interpolates between a regime in which different measurement
outcomes remain entanglement-distinguishable and one in which their universal
entropies become asymptotically indistinguishable.  The corresponding
short-cylinder form and the derivation of both limits are given in
Appendix~\ref{app:limits}.

\section{Discussion}
We have shown that, in the scaling regime, the outcome-resolved R\'enyi
entanglement left after a local measurement of a critical quantum chain is
controlled by a single compact conformal coordinate extracted directly from
the measurement record.  At the XX point this relation is connected to exact
free-fermion identities and an analytical asymptotic matching to the compact
boundary mode, while the compact-boson description identifies the surviving
infrared degree of freedom.  The interacting XXZ results show that the same
microscopic coordinate continues to organize the conditional entropy across
the critical Luttinger-liquid regime without interaction-dependent retuning.
Its agreement across subsystem geometries, R\'enyi indices, interactions, and
system sizes points to a common infrared mechanism in which the universal
entropy depends only on the relative compact boundary mode selected by the
conformal geometry. This reduction is nevertheless observable-specific.  Fixing $\delta_L$ does
not determine the full conditioned state: local observables, correlation
matrices, and the entanglement Hamiltonian may retain additional outcome
dependence \cite{EislerTonni2026}.  What becomes universal here is the
outcome-resolved R\'enyi entropy.  The compact-boson cancellation holds for
arbitrary coupling, whereas at the XX point the microscopic problem admits
exact free-fermion identities and an analytical asymptotic matching to the
compact boundary mode.  For interacting XXZ chains, the numerical results
identify the same conformally weighted snapshot coordinate as the relevant
microscopic variable, but a corresponding analytical derivation remains open. More broadly, these results show that predicting a universal property of a
postmeasurement state need not require reconstruction of the full conditioned
wave function.  An exponentially large measurement record can retain
substantial microscopic information while its long-distance entanglement
depends on only one geometrically selected degree of freedom.  In this sense
the compression is extreme: the conditioned state remembers the snapshot,
while its universal entanglement remembers only the compact conformal
coordinate.

\begin{acknowledgments}
M.A.R. acknowledges partial support from CNPq and FAPERJ (Grant
No.~E26/210.062/2023). We thank M. Heyl, E. Tonni, and V. Eisler for discussions. We gratefully acknowledge the resources on the LiCCA HPC cluster of the University of Augsburg, co-funded by the Deutsche Forschungsgemeinschaft (DFG, German Research Foundation)–Project-ID 499211671.
\end{acknowledgments}

%\paragraph*{Data availability.—}
%The numerical data and custom code supporting this study are available from the
%author upon reasonable request and will be deposited in a persistent public
%repository before publication.

\clearpage
\onecolumngrid

\appendix

\section{Model, measurement geometry, and microscopic coordinate}
\label{sec:model}

This section fixes the microscopic conventions, measurement geometry, and the
two quantities compared throughout the paper: the conditional entanglement
entropy and the compact coordinate extracted directly from the measurement
record. For a given outcome $m$, these are constructed independently. At the
XX point the conditioned state admits exact free-fermion formulas, whereas for
interacting XXZ chains the corresponding quantities are evaluated by exact
diagonalization or periodic matrix-product states. The limiting XX
distribution of the compact coordinate is taken from
Ref.~\cite{RajabpourSnapshots2026}.

\subsection{Microscopic chains and partial measurement}

We consider the periodic spin-$1/2$ XXZ chain
\begin{equation}
 H(\Delta)=\sum_{j=0}^{L-1}\left(
 S_j^xS_{j+1}^x+S_j^yS_{j+1}^y+\Delta S_j^zS_{j+1}^z
 \right),
 \qquad S_L\equiv S_0 ,
 \label{eq:xxz}
\end{equation}
at zero magnetization and even $L$. The critical regime is
$-1<\Delta\le1$, with the XX point at $\Delta=0$. In the compact-boson
normalization used below, the exact Luttinger coupling is
\begin{equation}
 g(\Delta)=\frac{\arccos(-\Delta)}{\pi},
 \qquad
 K=\frac{1}{2g(\Delta)},
 \label{eq:gdelta}
\end{equation}
so that $g(0)=1/2$ and $K=1$
\cite{Haldane1981a,Haldane1981b,Giamarchi2003}.

At $\Delta=0$, the Jordan--Wigner transformation maps the model to free
fermions, and the half-filled ground state is a Slater determinant of
$N=L/2$ consecutive momenta
\cite{JordanWigner1928,LiebSchultzMattis1961,Katsura1962}. In the occupation
basis, the probability of a complete configuration $X$ is
\begin{equation}
 p_L(X)=L^{-N}\prod_{x<y\in X}
 \left|\e^{2\pi\ii x/L}-\e^{2\pi\ii y/L}\right|^2,
 \qquad |X|=N .
 \label{eq:born}
\end{equation}
Thus the occupations form a rank-$N$ projection determinantal process,
equivalently a discrete circular $\beta=2$ log gas
\cite{Macchi1975,Soshnikov2000,Johansson1998}. This microscopic probability
law also places the XX chain in the standard framework of free-fermion
counting statistics
\cite{AbanovIvanov2011,IvanovLevkivskyi2016,CostinLebowitz1995,
Soshnikov2002,DiaconisEvans2001,LevitovLesovik1993,Klich2003,
KlichLevitov2009,Song2012,CalabreseMintchevVicari2012,Gamayun2020}.

For all $\Delta$ we use the same measured binary variable
$m_j=S_j^z+1/2\in\{0,1\}$. Four consecutive arcs are ordered as
$A,C_1,B,C_2$ around the ring, and only $C=C_1\cup C_2$ is measured in the
occupation/$S^z$ basis. A measurement record is
$m=(m_j)_{j\in C}$. We denote the normalized conditional state on
$U=A\cup B$ by $\ket{\Psi_m}$ and its reduced density matrix on $A$ by
$\rho_A(m)$. For real $\alpha>0$, $\alpha\neq1$, the outcome-resolved
R\'enyi entropy is
\begin{equation}
 S_{\alpha,L}(m)=\frac{1}{1-\alpha}
 \log\Tr\rho_A(m)^\alpha .
 \label{eq:renyi}
\end{equation}
The von Neumann limit is discussed in Sec.~\ref{sec:consequences}. The exact
microscopic derivation developed below specializes to $\Delta=0$, while the
compact-boson calculation is kept at general $g$ and the interacting XXZ
results test the same lattice coordinate away from the free-fermion point.

\subsection{Conformal geometry and record coordinate}

Let $z_a=\exp(2\pi\ii x_a/L)$, $a=1,\ldots,4$, denote the ordered endpoints of
the four arcs. Their conformal cross ratio is
\begin{equation}
 \zeta=
 \left|
 \frac{(z_1-z_2)(z_3-z_4)}
 {(z_1-z_3)(z_2-z_4)}
 \right|
 \in(0,1),
 \label{eq:cross-ratio}
\end{equation}
and the corresponding cylinder has circumference $2\pi$ and height
\begin{equation}
 h(\zeta)=\pi\frac{K_{\rm ell}(\zeta)}
 {K_{\rm ell}(1-\zeta)},
 \label{eq:h-zeta}
\end{equation}
where $K_{\rm ell}$ is the complete elliptic integral of the first kind
\cite{Rajabpour2015,Rajabpour2016,Stephan2014}. Thus $h$ is fixed entirely
by the interval geometry. If $W=u+iv$ maps the four marked endpoints to the
corresponding rectangle, the harmonic side coordinate
$f=2\,\mathrm{Im}\,W$ is constant on the two unmeasured arcs and winds once,
with opposite orientations, along the two measured arcs. Sampling it at
lattice midpoints gives the geometric weights $f_j$.

The compact coordinate introduced in Ref.~\cite{RajabpourSnapshots2026} is
\begin{equation}
 \delta_L(m)=
 \left[
 \sum_{j\in C}f_j\left(m_j-\frac12\right)-\pi|B|
 \right]_{2\pi},
 \label{eq:delta}
\end{equation}
where $[x]_{2\pi}$ denotes $x$ modulo $2\pi$, represented in $[0,2\pi)$.
The coordinate is not a simple particle count or imbalance: the harmonic
weights are fixed by the four endpoints and select a specific long-wavelength
component of the measured record. It depends only on the measured bits; the
unknown occupation of $B$ can change a real lift only by an integer multiple
of the compact period.

At the XX point, the scaling-limit Born density of $\delta_L$ is the heat
kernel on the circle derived in Ref.~\cite{RajabpourSnapshots2026},
\begin{equation}
 P_h(\delta)=\frac{1}{2\pi}\sum_{p\in\bbZ}
 \e^{-hp^2-\ii p\delta}.
 \label{eq:delta-law}
\end{equation}
This distribution will be used only in Sec.~\ref{sec:consequences}; until
then, all statements are conditioned on a fixed measurement record.

\subsection{Exact conditional correlation matrix}

At the XX point let $G$ be the ground-state Fermi projector,
$M_m=\mathrm{diag}(m_j)_{j\in C}$, and
$H_m=G_{CC}+M_m-I_C$, where $I_C$ is the identity on the measured sites.
Conditioning on the occupations in $C$ leaves a Slater determinant on
$U=A\cup B$ with correlation projector
\begin{equation}
 G_U^{(m)}
 =
 G_{UU}-G_{UC}H_m^{-1}G_{CU}.
 \label{eq:conditional}
\end{equation}
If $\nu_\beta\in[0,1]$ are the eigenvalues of its restriction to $A$, the
Gaussian correlation-matrix formula gives
\begin{equation}
 S_{\alpha,L}(m)
 =
 \frac{1}{1-\alpha}
 \sum_\beta
 \log\!\left[
 \nu_\beta^\alpha+(1-\nu_\beta)^\alpha
 \right].
 \label{eq:spectrum-entropy}
\end{equation}
These equations provide the exact finite-size entropy used independently in
the XX numerical tests \cite{Peschel2003}. The corresponding conditioning
and determinant identities are derived in Sec.~\ref{app:gaussian}.

At this stage the measurement record and the entropy have been defined by
independent constructions. The next sections determine how the complete
microscopic record enters the replica problem and how its long-distance
dependence reduces to the compact coordinate $\delta_L(m)$.

\section{Compact boson with an arbitrary boundary profile}
\label{sec:cft}

We next isolate the continuum mechanism underlying the outcome-resolved
decoder.  Consider a compact boson $\phi\equiv\phi+2\pi$ on a cylinder
$[0,h]\times S^1$ with action
\begin{equation}
 S[\phi]
 =
 \frac{g}{4\pi}
 \int_0^h\!\dd u\int_0^{2\pi}\!\dd v\,|\nabla\phi|^2 .
 \label{eq:boson-action}
\end{equation}
We impose arbitrary periodic Dirichlet profiles on the two boundaries,
\begin{equation}
 \varphi_a(v)
 =
 \varphi_{a,0}
 +
 \sum_{k\neq0}\varphi_{a,k}\e^{\ii kv},
 \qquad a=1,2,
 \label{eq:profile-fourier}
\end{equation}
with no tangential winding, and define their relative compact average by
$\delta=[\varphi_{2,0}-\varphi_{1,0}]_{2\pi}$.

For fixed boundary profiles, the Gaussian Dirichlet problem separates into
the oscillator contribution, the nonzero Fourier modes of the boundary data,
and the compact zero-mode sector.  Up to a profile-independent normalization,
the cylinder partition function is
\begin{align}
 Z_h[\varphi_1,\varphi_2]
 &=
 \eta(q_1)^{-1}
 \e^{-S_{{\rm nz},h}[\varphi_1,\varphi_2]}
 \Theta_1^{(g)}(h,\delta),
 \label{eq:profile-Z}\\
 q_1&=\e^{-2\pi^2/h},
 \nonumber
\end{align}
where $S_{{\rm nz},h}$ contains all nonzero boundary modes and
\begin{equation}
 \Theta_n^{(g)}(h,\delta)
 =
 \sum_{w\in\bbZ}
 \exp\!\left[
 -\frac{ng}{2h}(\delta+2\pi w)^2
 \right].
 \label{eq:theta-g}
\end{equation}
The eta function is the homogeneous Dirichlet oscillator determinant, while
the theta sum runs over compact radial-winding sectors.  Related compact-boson
replica constructions were developed in
Refs.~\cite{FradkinMoore2006,HsuMulliganFradkinKim2009}.  In our convention
the XX point corresponds to $g=1/2$, whereas in the critical XXZ chain the
continuum coupling is fixed by Eq.~\eqref{eq:gdelta}.

\subsection{Replica cover and cancellation of nonzero modes}

On the connected $n$-sheet replica surface, uniformization maps the cylinder
height to $h/n$ and pulls back the boundary profiles as
$(P_n\varphi)(v)=\varphi(nv)$.  A Fourier mode $k$ is therefore mapped to
mode $nk$.  The nonzero-mode action obeys
\begin{equation}
 S_{{\rm nz},h/n}[P_n\varphi_1,P_n\varphi_2]
 =
 n\,S_{{\rm nz},h}[\varphi_1,\varphi_2].
 \label{eq:covering}
\end{equation}
This is exactly the factor appearing in the product of the $n$ one-sheet
normalizations.  Hence every nonzero Fourier component of the two boundary
profiles cancels from the normalized replica ratio before any Born average is
taken.

The compact zero mode behaves differently.  The connected replica surface
contains a single radial-winding sum, whereas the denominator contains $n$
independent winding sums.  The resulting fixed-profile ratio is therefore
\begin{equation}
 \cR_n^{(g)}(h,\delta)
 =
 \frac{\eta(q_1)^n}{\eta(q_n)}
 \frac{\Theta_n^{(g)}(h,\delta)}
      {\Theta_1^{(g)}(h,\delta)^n},
 \qquad
 q_n=\e^{-2\pi^2n/h},
 \label{eq:Rn}
\end{equation}
and the corresponding universal R\'enyi function is
\begin{equation}
 \cF_n^{(g)}(h,\delta)
 =
 \frac{1}{1-n}\log\cR_n^{(g)}(h,\delta).
 \label{eq:Fn-g}
\end{equation}
Thus, for an arbitrary periodic boundary profile, the normalized
outcome-resolved entropy is insensitive to every finite-frequency deformation
and retains only the relative compact average $\delta$.  This statement
concerns the entropy ratio, not the full conditioned state, and holds for
arbitrary compact-boson coupling $g$.

For comparison with lattice R\'enyi entropies at noninteger index, we use the
natural continuation $n\to\alpha$ of Eq.~\eqref{eq:Rn}.  This continuation
belongs only to the continuum prediction; the lattice entropies are evaluated
independently for each real $\alpha$.  The mode-by-mode Dirichlet calculation
and the treatment of the compact sector are given in
Sec.~\ref{app:boson}.

The continuum result by itself does not determine which microscopic variable
of an interacting lattice snapshot realizes the surviving zero mode.  At the
XX point that identification follows from the free-fermion construction
developed next, while for XXZ it is tested numerically without redefining the
lattice coordinate.

\section{Dirichlet-to-Neumann calculation for the compact boson}
\label{app:boson}

The previous section identifies the microscopic record with a compact boundary
profile whose relative zero mode is $\delta_L(m)$, while its nonzero Fourier
components retain the remaining profile dependence.  We now show explicitly
that these nonzero modes cancel from the normalized replica ratio.  The
calculation also gives the Dirichlet on-shell action used in
Sec.~\ref{sec:cft}.

Expand a real boundary profile as
\begin{equation}
 \varphi_a(v)=\varphi_{a,0}
 +\sum_{k\ne0}\varphi_{a,k}\e^{\ii kv},
 \qquad
 \varphi_{a,-k}=\varphi_{a,k}^*,
 \qquad a=1,2.
\end{equation}
For each pair of modes $\pm k$, with $k>0$, the harmonic extension of the
positive-frequency coefficient between the two boundaries is
\begin{equation}
 \phi_k(u)=
 \frac{\sinh[k(h-u)]}{\sinh(kh)}\,\varphi_{1,k}
 +\frac{\sinh(ku)}{\sinh(kh)}\,\varphi_{2,k}.
\end{equation}
Using Green's identity, the contribution of the pair $\pm k$ to the on-shell
action is
\begin{align}
 S_k
 =\frac{gk}{\sinh(kh)}
 \bigg[
 &\cosh(kh)
 \left(
 |\varphi_{1,k}|^2+|\varphi_{2,k}|^2
 \right)
 \notag\\
 &-2\,\mathrm{Re}
 \left(
 \varphi_{1,k}^*\varphi_{2,k}
 \right)
 \bigg].
\end{align}
Introducing the symmetric and antisymmetric combinations
\begin{equation}
 \varphi_{\pm,k}
 =
 \frac{\varphi_{2,k}\pm\varphi_{1,k}}{\sqrt2},
\end{equation}
and summing over $k>0$ gives the complete nonzero-mode contribution
\begin{equation}
 S_{{\rm nz},h}
 =
 g\sum_{k\ge1}k
 \left[
 \tanh\!\frac{kh}{2}\,|\varphi_{+,k}|^2
 +
 \coth\!\frac{kh}{2}\,|\varphi_{-,k}|^2
 \right].
 \label{eq:app-nz}
\end{equation}

The zero mode is treated separately.  Its harmonic solution is linear in the
cylinder coordinate $u$, and compactness allows the relative boundary value to
differ by an integer winding,
\begin{equation}
 \varphi_{2,0}-\varphi_{1,0}
 \longrightarrow
 \delta+2\pi w,
 \qquad w\in\bbZ.
\end{equation}
The corresponding classical action is
\begin{equation}
 S_{0,w}
 =
 \frac{g}{2h}
 (\delta+2\pi w)^2.
 \label{eq:app-winding}
\end{equation}
Gaussian fluctuations around the classical solution obey homogeneous
Dirichlet boundary conditions and contribute the standard cylinder determinant
$\eta(q_1)^{-1}$.

We now consider the connected $n$-sheet replica surface.  Uniformization maps
it to a cylinder of height $h/n$, while the physical boundary profile is pulled
back according to
\begin{equation}
 (P_n\varphi)(v)=\varphi(nv).
\end{equation}
Thus a Fourier mode $k$ on the original cylinder becomes mode $nk$ on the
cover,
\begin{equation}
 (P_n\varphi)_{nk}=\varphi_k.
\end{equation}
Applying Eq.~\eqref{eq:app-nz} at height $h/n$ gives, mode by mode,
\begin{align}
 &g(nk)
 \tanh\!\left[
 \frac{(nk)(h/n)}{2}
 \right]
 |\varphi_{+,k}|^2
 =
 n\,gk
 \tanh\!\frac{kh}{2}
 |\varphi_{+,k}|^2,
 \\
 &g(nk)
 \coth\!\left[
 \frac{(nk)(h/n)}{2}
 \right]
 |\varphi_{-,k}|^2
 =
 n\,gk
 \coth\!\frac{kh}{2}
 |\varphi_{-,k}|^2.
\end{align}
Therefore
\begin{equation}
 S_{{\rm nz},h/n}
 [P_n\varphi_1,P_n\varphi_2]
 =
 n\,
 S_{{\rm nz},h}
 [\varphi_1,\varphi_2],
 \label{eq:covering}
\end{equation}
which is precisely the factor required for cancellation against the $n$
one-sheet normalization factors.  Hence every nonzero Fourier component of an
arbitrary fixed boundary profile cancels exactly from the normalized replica
ratio.

The relative compact winding is different.  The connected replica surface
contains a single winding sum, whereas the denominator contains $n$
independent one-sheet sums.  Combining this zero-mode contribution with
Eq.~\eqref{eq:app-nz} and the Dirichlet fluctuation determinant reproduces the
fixed-profile partition function and replica ratio of
Eqs.~\eqref{eq:profile-Z} and \eqref{eq:Rn}.  Thus the normalized entropy
forgets the entire nonzero-mode boundary profile and retains only the relative
compact zero mode $\delta$.

\section{Exact outcome-resolved replica representation at the XX point}
\label{sec:lattice}

We now return to the microscopic XX chain and make explicit how a fixed
measurement record enters the R\'enyi moment.  Because conditioning preserves
Gaussianity, the problem admits an exact finite-size representation in terms
of charge-counting functions.  For integer R\'enyi index this representation
can then be rewritten as a discrete two-cut Coulomb gas.  These steps retain
the complete binary record and involve neither Born averaging nor a smooth
boundary approximation.

\subsection{Counting function and replica phases}

Let $\Pi_m$ project onto the measured occupation record $m$, and let
$N_A=\sum_{j\in A}c_j^\dagger c_j$ count fermions in $A$.  Define the
unnormalized outcome-resolved counting function
\begin{equation}
 D_{m,L}(\lambda)
 =
 \bra{\Psi_0}\Pi_m\lambda^{N_A}\ket{\Psi_0},
 \qquad
 D_{m,L}(1)=p_L(m).
 \label{eq:counting}
\end{equation}
Thus $D_{m,L}(\lambda)/p_L(m)$ is the particle-number generating function in
the state conditioned on the same record $m$.  For integer $n\ge2$, introduce
the phases
\begin{equation}
 \lambda_r=
 \exp\!\left[\frac{\ii\pi}{n}(2r+1-n)\right],
 \qquad
 r=0,\ldots,n-1 .
 \label{eq:phases}
\end{equation}
Since
\begin{equation}
 \prod_{r=0}^{n-1}
 \bigl(1-\nu+\lambda_r\nu\bigr)
 =
 (1-\nu)^n+\nu^n ,
 \label{eq:phase-id}
\end{equation}
the conditional R\'enyi moment factorizes exactly as
\begin{equation}
 \Tr\rho_A(m)^n
 =
 \prod_{r=0}^{n-1}
 \frac{D_{m,L}(\lambda_r)}{p_L(m)} .
 \label{eq:replica-factor}
\end{equation}
Equation~\eqref{eq:replica-factor} is the outcome-resolved form of the
standard fermionic full-counting-statistics identity
\cite{Klich2003}.  Every factor refers to the same microscopic measurement
record, so the reduction to a smaller set of infrared variables cannot arise
from averaging over outcomes.  The equivalent determinant formulas are
collected in Sec.~\ref{app:gaussian}.

\subsection{Pinned two-cut Coulomb gas}

Let $O_m\subset C$ be the set of measured occupied sites and
$N_m^U=N-|O_m|$ the number of particles remaining on
$U=A\cup B$.  Expanding Eq.~\eqref{eq:counting} in the position basis gives
\begin{align}
 D_{m,L}(\lambda)
 &=
 L^{-N}
 \sum_{\substack{Y\subset A\cup B\\ |Y|=N_m^U}}
 |\Delta(z_{O_m\cup Y})|^2\lambda^{|Y\cap A|},
 \label{eq:pinned}\\
 z_x&=\e^{2\pi\ii x/L},
 \nonumber
\end{align}
where $\Delta(z_S)=\prod_{x<y\in S}(z_x-z_y)$ is the Vandermonde determinant.
Measured occupied sites therefore act as pinned logarithmic charges, measured
empty sites are excluded, and the remaining particles form a discrete
two-cut Coulomb gas on $A\cup B$.  The factor $\lambda$ couples to the
relative filling of the two unmeasured cuts.

The full microscopic pattern remains in the external potential generated by
the pinned charges.  In particular, a typical binary outcome is not replaced
at this stage by a smooth or uniform boundary condition.  The collective
long-wavelength variable relevant to the two-cut problem is the transfer of
an integer number of particles between $A$ and $B$; the next step identifies
the corresponding displacement with the compact coordinate extracted from
the measured record.

\subsection{Boundary height and the relative compact mode}

For a complete occupation configuration, let $n_j\in\{0,1\}$ denote the
occupation of site $j$, with $n_j=m_j$ fixed on the measured region.  Introduce
the compact lattice height through
\begin{equation}
 \Phi_{j+1}-\Phi_j
 =
 2\pi\left(n_j-\frac12\right).
 \label{eq:height}
\end{equation}
For the circular log gas, the long-wavelength height and logarithmic potential
are harmonic conjugates.  Ref.~\cite{RajabpourSnapshots2026} identifies the
coarse-grained record with compact boundary profiles
$\varphi_a^{[L]}(v;m)$, $a=1,2$, on the two sides of the conformal cylinder.
Writing their angular averages as
$\varphi_{a,0}^{[L]}=(2\pi)^{-1}\int_0^{2\pi}
\varphi_a^{[L]}(v;m)\,\dd v$, the microscopic coordinate becomes
\begin{equation}
 \delta_L(m)
 =
 \big[
 \varphi_{2,0}^{[L]}(m)-\varphi_{1,0}^{[L]}(m)
 \big]_{2\pi}.
 \label{eq:zero-mode}
\end{equation}
Thus the conformally weighted bit-string coordinate is the relative compact
boundary average in the scaling description.  The harmonic-coordinate
derivation of Eq.~\eqref{eq:zero-mode} is given in
Sec.~\ref{app:height}.

At this point the nonzero Fourier components of the rough record-dependent
profile are still present in the microscopic potential.  The continuum result
of Sec.~\ref{sec:cft} shows that such finite-frequency components cancel from
the normalized replica ratio once a regulated boundary profile is reached.
The remaining task is therefore microscopic: to connect the discrete
two-cut filling sectors to that compact zero mode and to control the
lattice-scale roughness of a binary snapshot.  This matching is developed in
the next section.

\section{Conditional Slater determinants and counting identities}
\label{app:gaussian}

For completeness, we collect here the exact finite-size free-fermion identities
underlying the XX construction.  These identities provide both the conditioned
correlation matrix used in the DPP calculation of
Sec.~\ref{sec:xx-numerics} and the counting-polynomial representation entering
the replica formulation.

For a Slater determinant with correlation matrix $G$ and a set $C$ of measured
sites, the probability of a binary record $m$ is
\begin{align}
 p_L(m)&=(-1)^{|C|-N_m}\det H_m,
 \label{eq:app-prob}\\
 H_m&=G_{CC}+M_m-I_C,
\end{align}
where $N_m=\sum_{j\in C}m_j$.  Equation~\eqref{eq:app-prob} follows directly
from Wick contractions, or equivalently from the principal-minor expansion of
the standard fermionic occupation generating determinant
\cite{NajafiRajabpourPRB2016}.

Projective measurements of individual occupations preserve Gaussianity and
update the residual correlation matrix by the rank-one conditioning rules used
in Sec.~\ref{sec:xx-numerics} \cite{Cheng2024}.  Iterating these updates over
the complete measurement record and performing block Gaussian elimination gives
the equivalent compact multi-site expression Eq.~\eqref{eq:conditional}.
Since the postmeasurement state remains a Slater determinant, the restriction
to $A$ has single-particle eigenvalues $\nu_\beta$, from which
Eq.~\eqref{eq:spectrum-entropy} follows.

For the counting polynomial one may apply the standard free-fermion
full-counting-statistics determinant identity \cite{Klich2003}.  Equivalently,
once the record is conditioned, the normalized generating function is
determined only by the $A$ block of the conditional correlation matrix:
\begin{equation}
 \frac{D_{m,L}(\lambda)}{p_L(m)}
 =\det\big[I_A+(\lambda-1)G_A^{(m)}\big]
 =\prod_\beta(1-\nu_\beta+\lambda\nu_\beta).
 \label{eq:app-fcs}
\end{equation}
Multiplying Eq.~\eqref{eq:app-fcs} over the phases
Eq.~\eqref{eq:phases} immediately gives Eq.~\eqref{eq:replica-factor}.  Exact
$\nu=0$ or $1$ modes, when present because of rank constraints, contribute
trivially and should simply be omitted from constructions that divide by
$\nu(1-\nu)$.

These identities establish the exact finite-size route from a measurement
record to its conditional entropy and replica observable.  We next turn to the
complementary question of how the same record enters the pinned-particle
representation and reduces, in the scaling description, to the compact
coordinate $\delta_L(m)$.

\section{From the pinned potential to the compact boundary average}
\label{app:height}

The harmonic-coordinate argument connecting the exact pinned-particle
representation to the compact boundary coordinate of
Ref.~\cite{RajabpourSnapshots2026} is summarized here.  It explains why the
harmonic, rather than uniform, weighting of the measured record is selected and
yields Eq.~\eqref{eq:zero-mode}.

Writing the Vandermonde in Eq.~\eqref{eq:pinned} as pinned--pinned,
pinned--mobile, and mobile--mobile factors gives a two-cut
orthogonal-polynomial ensemble with weight
\begin{equation}
 w_m(x)=\mathbf 1_{A\cup B}(x)
 \prod_{o\in O_m}|z_x-z_o|^2 .
\end{equation}
The logarithm of $w_m$ is the discrete electrostatic potential generated by
the occupied measured sites.  After subtracting the half-filled background,
the Fourier representation
\begin{equation}
 -\log|\e^{\ii\theta}-\e^{\ii\theta'}|
 =
 \sum_{k\ge1}
 \frac{\cos k(\theta-\theta')}{k}
 +\mathrm{const}
\end{equation}
shows that the long-wavelength logarithmic potential and the compact height
defined in Eq.~\eqref{eq:height} are harmonic conjugates.  Thus the same
microscopic record may be viewed either as a pinned-charge potential or as a
boundary-height profile.

Let $f_j$ be the midpoint sampling of the harmonic side coordinate introduced
in Sec.~\ref{sec:model}, and define the discrete difference
$\nabla f_j=f_j-f_{j-1}$.  Using Eq.~\eqref{eq:height}, discrete summation by
parts gives
\begin{equation}
 \sum_{j\in C}
 f_j\left(m_j-\frac12\right)
 =
 -\frac{1}{2\pi}
 \sum_{j\in C}\Phi_j\,\nabla f_j
 +\text{boundary terms}.
\end{equation}
In the scaling description the bulk term becomes the harmonic pairing of the
coarse-grained height with the conformal side coordinate.  The boundary terms
at the arc endpoints fix the choice of real lift; with the convention of
Eq.~\eqref{eq:delta}, they produce the $-\pi|B|$ shift.  Since $f$ is constant
on each unmeasured arc and winds with opposite orientations along the two
measured arcs, the resulting compact quantity is precisely the difference of
the two boundary averages,
\begin{equation}
 \delta_L(m)
 =
 \big[
 \varphi_{2,0}^{[L]}(m)
 -
 \varphi_{1,0}^{[L]}(m)
 \big]_{2\pi},
\end{equation}
which is Eq.~\eqref{eq:zero-mode}.

Only this relative average is invariant under adding a common compact shift to
the two boundaries.  The nonzero Fourier components of the same record remain
in the pinned-charge potential and are carried into the arbitrary-profile
continuum problem of Sec.~\ref{sec:cft}.  The next section shows explicitly
that these nonzero modes cancel from the normalized replica ratio, leaving only
the compact relative zero mode.

\section{Microscopic matching at the free-fermion point}
\label{sec:matching}

The previous two sections establish complementary pieces of the XX problem.
The compact-boson calculation shows that a regulated boundary profile affects
the normalized R\'enyi entropy only through its relative compact zero mode,
while the exact lattice representation retains the full binary measurement
record.  Connecting these descriptions requires two steps: first, matching a
fixed set of long-wavelength record modes to the two-cut compact-boson sector;
second, showing that the remaining lattice-scale roughness does not contribute
a finite correction to the normalized entropy.  This separation makes clear
which ingredients are exact and which enter as scaling inputs.

\subsection{Fixed long-wavelength cutoff}

Introduce a conformal momentum cutoff $\Lambda$ and retain only the smooth
record modes below that scale.  Let $Z_{L,\Lambda}(s;m)$ denote the
fixed-filling partition function in the sector obtained by transferring an
integer $s$ of particles between the two unmeasured cuts, and define
$a_L=\delta_L(m)/(2\pi)$ modulo integers.  For every fixed $s$, the required
large-$L$ asymptotic form is
\begin{equation}
 \log\frac{Z_{L,\Lambda}(s;m)}
 {Z_{L,\Lambda}(0;m)}
 =
 -\frac{\pi^2}{h}
 \big[(s+a_L)^2-a_L^2\big]+o(1).
 \label{eq:fixedK-ratio}
\end{equation}
Together with tightness of the filling-transfer sectors, this gives the
relative weights
\begin{equation}
 \mathcal W_L(s\mid m)
 \propto
 \exp\!\left[
 -\frac{\pi^2}{h}
 \left(s+\frac{\delta_L(m)}{2\pi}\right)^2
 \right],
 \qquad s\in\bbZ .
 \label{eq:discrete-gaussian}
\end{equation}
The coefficient $\pi^2/h$ is the genus-one filling stiffness of the two-cut
problem, while the center is fixed by the compact displacement
$\delta_L/(2\pi)$.

The detailed derivation is given in Sec.~\ref{app:matching}.  An exact
stereographic transformation maps the roots-of-unity gas to a discrete
Vandermonde-squared ensemble on an analytic quantile lattice with two hard
allowed blocks.  The equilibrium occupation of the available nodes is
$1/2$, so the discrete void and saturation constraints remain inactive.  The
quadratic stiffness is then fixed by the genus-one period matrix.  The center
is determined in two steps: an exact finite-lattice particle--hole symmetry
fixes the neutral integer/half-integer characteristic, and the smooth
record-dependent harmonic response shifts it by the measured conformal phase.
Hence neither the stiffness nor the center is a fitting parameter.

What remains at fixed $\Lambda$ is a standard local fixed-filling asymptotic
input, uniform for $O(1)$ filling transfers and retained profile amplitudes in
compact sets, together with tightness of the filling sectors.  Existing
continuous multi-cut, discrete $\beta$-ensemble, and discrete
orthogonal-polynomial results establish the relevant ingredients separately
\cite{BorotGuionnet2024,BorodinGorinGuionnet2017,
BorotGorinGuionnet2026,BaikKriecherbauerMcLaughlinMiller2003}, although the
precise combination of two hard blocks on the present analytic quantile
lattice with a free filling fraction is not covered directly.  Under this
input, summing the filling sectors reconstructs the compact theta function,
while the smooth nonzero-mode factors cancel by Eq.~\eqref{eq:covering}.

\subsection{Removing lattice-scale boundary roughness}

A microscopic binary record contains Fourier components up to the lattice
scale, particularly near the slit endpoints.  Let
$S_{n,L}^{(\Lambda)}(m)$ denote the entropy obtained after retaining only
record modes below the cutoff.  The exact finite-size response formula derived
in Sec.~\ref{app:matching} expresses
$S_{n,L}(m)-S_{n,L}^{(\Lambda)}(m)$ directly in terms of the omitted
high-frequency potential and the response of the conditioned kernel.  Thus no
pointwise smoothness assumption on the microscopic snapshot is required.

The remaining ultraviolet condition is that the cutoff can be increased with
system size while staying below the lattice scale,
$1\ll\Lambda_L\ll L$, such that
\begin{equation}
 \bbE_{\rm Born}
 \left|
 S_{n,L}(m)-S_{n,L}^{(\Lambda_L)}(m)
 \right|
 \longrightarrow0 .
 \label{eq:uv-assumption}
\end{equation}
If the fixed-cutoff asymptotics remain uniform along the same sequence
$\Lambda_L$, the microscopic decoder follows in probability:
\begin{equation}
 S_{n,L}(m)
 -
 \cF_n\!\left(h,\delta_L(m)\right)
 \xrightarrow[L\to\infty]{\mathbb P_{\rm Born}}0 .
 \label{eq:main-result}
\end{equation}

This ultraviolet issue is distinct from boundary locking in
R\'enyi-weighted formation probabilities, where the measurement
configurations themselves are reweighted and the effective boundary condition
may change \cite{StephanMisguichPasquier2011,
RajabpourShannonBoundary2026}.  Here the measurement record remains distributed
with its ordinary Born weight, and the replica index acts only on
$\Tr\rho_A(m)^n$.

The numerical tests below probe the combined consequence of these two
microscopic inputs.  If an additional record-dependent variable survived the
ultraviolet limit, outcomes with the same $\delta_L$ would retain a finite
entropy spread as the system size increases.

\section{Fixed-cutoff asymptotics and ultraviolet response}
\label{app:matching}

This section supplies the two microscopic ingredients required in
Sec.~\ref{sec:matching}.  We first analyze the problem at fixed conformal
cutoff $\Lambda$: an exact stereographic transformation reduces the circular
gas to a two-cut discrete orthogonal-polynomial ensemble, after which the
filling stiffness and theta characteristic can be fixed explicitly.  The only
remaining fixed-$\Lambda$ input is a local fixed-filling expansion with uniform
remainder and filling-sector tightness.  We then return to the full microscopic
record and derive an exact finite-size response formula for the Fourier modes
omitted by the cutoff.

\subsection{Exact reduction and strict-liquid structure}

Because the ensemble lives on roots of unity, its continuum density obeys
\begin{equation}
 0\le \frac{\dd\mu}{\dd\sigma}\le1,
 \qquad
 \dd\sigma=\frac{\dd\theta}{2\pi}.
\end{equation}
For a Born-typical half-filled record, adding the fixed measured background to
the mobile measure reduces the leading logarithmic energy to that of a
half-filled circular gas.  At leading logarithmic order the equilibrium measure
is therefore uniform, so the mobile part on the two unmeasured arcs is
\begin{equation}
 \dd\mu_\star
 =
 \frac12\,\dd\sigma\big|_{A\cup B}.
 \label{eq:app-liquid}
\end{equation}
Hence the equilibrium occupation of the available nodes is exactly $1/2$ and
lies a finite distance from both the void and saturation constraints.  The
fixed-$\Lambda$ record potential and the counting twist are subleading
perturbations and do not modify this macroscopic equilibrium density.

The circular geometry can be reduced exactly to a real-line problem.  Choose a
point $\theta_\star+\pi$ strictly inside one of the measured gaps and introduce
the stereographic coordinate
\begin{equation}
 x=\tan\!\frac{\theta-\theta_\star}{2}.
 \label{eq:stereo-map}
\end{equation}
The two unmeasured arcs then map to two compact real intervals.  For any two
lattice points,
\begin{equation}
 |\e^{\ii\theta_i}-\e^{\ii\theta_j}|^2
 =
 \frac{4(x_i-x_j)^2}
 {(1+x_i^2)(1+x_j^2)}.
 \label{eq:stereo-chord}
\end{equation}
If
\begin{equation}
 M=N-|O_m|
\end{equation}
particles remain mobile, then
\begin{equation}
 |\Delta(z_Y)|^2
 =
 4^{M(M-1)/2}\Delta(x_Y)^2
 \prod_{y\in Y}(1+x_y^2)^{-(M-1)}.
 \label{eq:stereo-vandermonde}
\end{equation}
The pinned--mobile factors transform in the same way and are absorbed into the
one-body weight.  Before the $\Lambda$ truncation, this weight may be written,
up to a record-dependent constant, as
\begin{equation}
 \widetilde w_m(x)
 \propto
 (1+x^2)^{-(N-1)}
 \prod_{o\in O_m}(x-x_o)^2.
 \label{eq:stereo-weight}
\end{equation}
The fixed-$\Lambda$ ensemble replaces the microscopic factor by its smooth
long-wavelength truncation.  Thus the regulated problem is an ordinary
Vandermonde-squared discrete orthogonal-polynomial ensemble on the nodes
\begin{equation}
 x_j
 =
 \tan\!\left[
 \frac{2\pi j/L-\theta_\star}{2}
 \right],
\end{equation}
restricted to two hard allowed intervals.  The normalized node-counting
measure has the explicit limit
\begin{equation}
 \frac1L\sum_j\delta_{x_j}
 \Longrightarrow
 \rho_0(x)\,\dd x,
 \qquad
 \rho_0(x)=\frac{1}{\pi(1+x^2)}.
 \label{eq:node-density}
\end{equation}
Equation~\eqref{eq:app-liquid} therefore becomes
\begin{equation}
 \dd\mu_\star
 =
 \frac{\rho_0(x)}{2}\,\dd x
\end{equation}
on both allowed intervals.  The equilibrium occupancy of the available nodes
is again exactly $1/2$, so the discrete upper and lower constraints are
inactive in a neighborhood of the equilibrium filling.  General discrete
orthogonal-polynomial asymptotics allow analytic non-equally-spaced node sets
of this type \cite{BaikKriecherbauerMcLaughlinMiller2003}; the additional
feature required here is the combination of two disconnected hard blocks with
a fluctuating filling fraction.

\subsection{Local fixed-filling input}

The fixed-$\Lambda$ analysis requires one local asymptotic statement.  Let
\begin{equation}
 M_L=N-|O_m|,
 \qquad
 \epsilon=\frac{\ell}{M_L},
\end{equation}
where $\ell$ is the number of mobile particles occupying one cut and
$\epsilon_\star$ is its equilibrium value.  Denote by
$Z_{L,\Lambda}(\epsilon;m)$ the corresponding fixed-filling partition
function.  For every fixed $\Lambda$ we require the standard off-critical
expansion
\begin{align}
 \log Z_{L,\Lambda}(\epsilon;m)
 &=
 M_L^2F_0(\epsilon)
 +M_LF_{1,\Lambda}(\epsilon;m)
 \notag\\
 &\quad
 +F_{2,\Lambda}(\epsilon;m)
 +o(1),
 \label{eq:fixed-filling-expansion}
\end{align}
uniformly for
\begin{equation}
 \epsilon=\epsilon_\star+\frac{s}{M_L}
\end{equation}
with $s$ in every fixed finite set, and uniformly over the retained
$|k|\le\Lambda$ profile amplitudes in a fixed compact set.  The leading
function $F_0$ is record independent.

We also require tightness of the associated filling-transfer distributions.
After normalizing the sector weights at fixed record, the total probability of
$|s|>S$ must tend to zero uniformly in $L$ as $S\to\infty$ over the same
compact family.  Local uniformity together with this tightness, rather than a
pointwise Gaussian bound, is sufficient to reconstruct the full filling-sector
sum.

Continuous multi-cut theory proves the corresponding fixed-filling expansion
and theta reconstruction \cite{BorotGuionnet2024}.  Discrete
$\beta$-ensemble theory establishes large deviations and discrete-Gaussian
filling fluctuations, including fixed, free, and constrained varying fillings
in broad multi-group settings
\cite{BorodinGorinGuionnet2017,BorotGorinGuionnet2026}.  General discrete
orthogonal-polynomial theory directly allows the analytic nonuniform quantile
lattice above \cite{BaikKriecherbauerMcLaughlinMiller2003}.  These results
establish the relevant ingredients separately, but they do not directly cover
the precise combination of two hard allowed blocks on the present analytic
quantile lattice with a free filling fraction.  We therefore use
Eq.~\eqref{eq:fixed-filling-expansion} together with filling-sector tightness
as the sole asymptotic input at fixed $\Lambda$.

\subsection{Filling stiffness from the genus-one mode}

Once the local expansion exists, its quadratic filling dependence is fixed.
Because Eq.~\eqref{eq:app-liquid} lies strictly inside the discrete constraint,
the constraint remains inactive for $O(M_L^{-1})$ filling changes.  The leading
function $F_0$ is therefore the ordinary two-cut logarithmic equilibrium energy
in that neighborhood.

Write
\begin{equation}
 \epsilon
 =
 \epsilon_\star+\frac{s}{M_L},
\end{equation}
with fixed integer $s$.  Since $F_0'(\epsilon_\star)=0$,
\begin{equation}
 \log\frac{Z_{L,\Lambda}(s;m)}
 {Z_{L,\Lambda}(0;m)}
 =
 \frac12F_0''(\epsilon_\star)s^2
 +\text{linear term}
 +o(1).
 \label{eq:filling-taylor}
\end{equation}
In the standard multi-cut expansion,
$F_0''/(2\pi\ii)$ is the filling period matrix; for $\beta=2$ it is the
period matrix of the genus-one spectral curve associated with the two cuts
\cite{BorotGuionnet2024}.  In the cycle convention adapted to particle
transfer, the elliptic uniformization of the four endpoints gives
\begin{equation}
 \tau_{\rm fill}
 =
 \ii\frac{K_{\rm ell}(1-\zeta)}
 {K_{\rm ell}(\zeta)}
 =
 \frac{\ii\pi}{h},
 \label{eq:filling-period}
\end{equation}
and hence
\begin{equation}
 F_0''(\epsilon_\star)
 =
 2\pi\ii\,\tau_{\rm fill}
 =
 -\frac{2\pi^2}{h}.
 \label{eq:filling-hessian}
\end{equation}
Equivalently, the same coefficient follows from the Dirichlet energy of the
condenser mode
\begin{equation}
 \chi(u,v)=\frac{u}{h},
\end{equation}
for which
\begin{equation}
 \int_{\mathcal C_h}|\nabla\chi|^2\,\dd^2x
 =
 \frac{2\pi}{h}.
\end{equation}
Thus the quadratic filling penalty is fixed to $\pi^2s^2/h$, independently of
the measurement record.

\subsection{Neutral characteristic and record-dependent displacement}

The center of the discrete Gaussian contains both a finite-lattice parity
characteristic and the smooth record response.  The neutral characteristic can
be fixed exactly.

Consider the half-filled reference in which the measured region is replaced by
its half-density background.  The corresponding mobile one-body weight is
\begin{equation}
 w_0(y)
 =
 \prod_{c\in C}|z_y-z_c|.
 \label{eq:neutral-weight}
\end{equation}
For any root of unity $z_y$,
\begin{equation}
 \prod_{\substack{x\in\Lambda_L\\x\neq y}}
 |z_y-z_x|
 =
 L.
 \label{eq:root-product}
\end{equation}
If $Y\subset D$ is the set of mobile particles and
$H=D\setminus Y$ the mobile holes, Eq.~\eqref{eq:root-product} gives the exact
identity
\begin{equation}
 |\Delta(z_Y)|^2
 \prod_{y\in Y}w_0(y)
 =
 \frac{L^{|Y|}}
 {\displaystyle
 \prod_{y\in Y,\,h\in H}|z_y-z_h|}.
 \label{eq:particle-hole-weight}
\end{equation}
At neutral half filling,
\begin{equation}
 |Y|=|H|=\frac{|U|}{2},
\end{equation}
so the weight is invariant under the particle--hole involution
$Y\leftrightarrow H$.  If
\begin{equation}
 \ell=|Y\cap A|,
\end{equation}
this involution sends
\begin{equation}
 \ell\longmapsto |A|-\ell
\end{equation}
and therefore fixes the center of the neutral filling distribution exactly at
$|A|/2$.  Relative to an integer reference filling, the neutral theta
characteristic is
\begin{equation}
 a_0
 =
 \frac{|A|}{2}\pmod{\bbZ}
 =
 \frac{|B|}{2}\pmod{\bbZ}.
 \label{eq:neutral-characteristic}
\end{equation}
Here we use the natural half-filled finite-size subsequence with $|U|$ even,
which is sufficient for the scaling limit considered here.

For the neutral record, Eq.~\eqref{eq:delta} gives
\begin{equation}
 \delta_0
 =
 [-\pi|B|]_{2\pi},
\end{equation}
and therefore
\begin{equation}
 a_0
 =
 \frac{\delta_0}{2\pi}
 \pmod{\bbZ}.
 \label{eq:neutral-delta}
\end{equation}
Thus the fixed $-\pi|B|$ term in the microscopic coordinate carries precisely
the finite-lattice integer/half-integer characteristic.

It remains to determine how a smooth record deformation moves this center.  At
fixed $\Lambda$, the retained one-body potential is smooth, so its first
variation under an $O(1)$ filling transfer is a deterministic continuum
electrostatic pairing; no fluctuating-sector asymptotic is required for this
step.  Let $\widetilde\delta_L$ and $\widetilde\delta_0$ be compatible real
lifts of the record and neutral boundary mismatches, and let
$\psi_\Lambda$ be the harmonic field generated by their difference.  With
$\chi(u,v)=u/h$,
\begin{equation}
 \int_{\mathcal C_h}
 \nabla\psi_\Lambda\!\cdot\!\nabla\chi\,\dd^2x
 =
 \frac{\widetilde\delta_L-\widetilde\delta_0}{h}
 \int_0^{2\pi}\dd v
 =
 \frac{2\pi(\widetilde\delta_L-\widetilde\delta_0)}{h}.
 \label{eq:record-zero-mode-pairing}
\end{equation}
All nonzero Fourier modes vanish in this pairing.  At the XX normalization
$g=1/2$, transferring $s$ compact units changes the record-dependent part of
the action by
\begin{equation}
 \Delta S_{\rm rec}(s)
 =
 \frac{\pi}{h}
 (\widetilde\delta_L-\widetilde\delta_0)s.
 \label{eq:record-linear-shift}
\end{equation}
Since the statistical weight is $\e^{-S}$, the neutral characteristic is
shifted to
\begin{equation}
 a_L
 =
 a_0
 +
 \frac{\widetilde\delta_L-\widetilde\delta_0}{2\pi}
 =
 \frac{\delta_L}{2\pi}
 \pmod{\bbZ}.
 \label{eq:theta-center}
\end{equation}
Combining Eqs.~\eqref{eq:filling-hessian} and
\eqref{eq:theta-center} gives, for every fixed $s$,
\begin{equation}
 \log\frac{Z_{L,\Lambda}(s;m)}
 {Z_{L,\Lambda}(0;m)}
 =
 -\frac{\pi^2}{h}
 \left[
 (s+a_L)^2-a_L^2
 \right]
 +o(1).
 \label{eq:app-multicut-ratio}
\end{equation}
Thus neither the stiffness nor the center of the filling-sector Gaussian is
left undetermined.  The fixed-$\Lambda$ input controls only the uniform
remainder and the filling-sector tails.

\subsection{From filling sectors to the compact theta function}

Set
\begin{equation}
 \mathfrak q
 =
 \e^{-\pi^2/h},
\end{equation}
and define the shifted counting theta function
\begin{equation}
 \vartheta_a(\lambda;\mathfrak q)
 =
 \sum_{s\in\bbZ}
 \mathfrak q^{(s+a)^2}\lambda^s.
 \label{eq:app-shifted-theta}
\end{equation}
Local uniformity in Eq.~\eqref{eq:app-multicut-ratio}, together with
filling-sector tightness, implies that the fixed-$\Lambda$ counting function is,
up to a nonuniversal reference filling and an $o(1)$ correction, a ratio of
shifted theta functions.  The reference filling drops out of the replica
product because
\begin{equation}
 \prod_{r=0}^{n-1}\lambda_r=1.
\end{equation}
The remaining product is fixed by the shifted Jacobi multiplication identity
\begin{equation}
 \prod_{r=0}^{n-1}
 \vartheta_a(\lambda_r;\mathfrak q)
 =
 \frac{\eta(\mathfrak q^2)^n}
 {\eta(\mathfrak q^{2n})}
 \vartheta_a(1;\mathfrak q^n),
 \label{eq:app-jacobi-multiplication}
\end{equation}
which follows from the Jacobi triple product and the fact that the
$\lambda_r$ are the roots of
\begin{equation}
 z^n-(-1)^{n+1}.
\end{equation}
Since
\begin{equation}
 q_1=\mathfrak q^2,
 \qquad
 q_n=\mathfrak q^{2n},
\end{equation}
and, for $g=1/2$,
\begin{equation}
 \vartheta_{a_L}(1;\mathfrak q^n)
 =
 \Theta_n(h,\delta_L),
\end{equation}
multiplying the replica phases gives
\begin{equation}
 \prod_{r=0}^{n-1}
 \frac{D_{m,L}^{(\Lambda)}(\lambda_r)}
 {D_{m,L}^{(\Lambda)}(1)}
 =
 \frac{\eta(q_1)^n}{\eta(q_n)}
 \frac{\Theta_n(h,\delta_L)}
 {\Theta_1(h,\delta_L)^n}
 +o(1).
 \label{eq:app-fixed-cutoff-replica-limit}
\end{equation}
Thus the discrete filling-transfer variable maps explicitly to the compact
radial winding sector.  At fixed $\Lambda$, every coefficient in the exponent
and the theta characteristic has been fixed; the only asymptotic input is the
local uniform expansion Eq.~\eqref{eq:fixed-filling-expansion} together with
filling-sector tightness for the two hard blocks of the analytic quantile
lattice.

\subsection{Exact ultraviolet response}

We now compare the fixed-$\Lambda$ problem with the full microscopic record.
Write the exact and truncated potentials as
\begin{equation}
 V
 =
 V^{(\Lambda)}
 +
 R^{(\Lambda)},
\end{equation}
and interpolate through
\begin{equation}
 V_t
 =
 V^{(\Lambda)}
 +
 tR^{(\Lambda)},
 \qquad
 0\le t\le1.
\end{equation}
Differentiating the logarithm of each finite-size counting determinant gives
\begin{equation}
 \frac{\dd}{\dd t}
 \log D_{m,L,t}(\lambda)
 =
 -\sum_{x\in A\cup B}
 R_{m,L}^{(\Lambda)}(x)
 \rho_{m,L,t}^{(\lambda)}(x),
\end{equation}
where $\rho_{m,L,t}^{(\lambda)}$ is the one-point function in the corresponding
tilted conditioned ensemble.  Combining the replica phases and subtracting the
one-sheet normalization gives the exact finite-size identity
\begin{align}
 (1-n)
 \big[
 S_{n,L}
 -
 S_{n,L}^{(\Lambda)}
 \big]
 &=
 -\int_0^1\!\dd t
 \sum_{x\in A\cup B}
 R_{m,L}^{(\Lambda)}(x)
 \notag\\[-2pt]
 &\hspace{2.5em}\times
 \mathcal J_{n,L,t}(x;m),
 \label{eq:app-uv-response}\\
 \mathcal J_{n,L,t}(x;m)
 &=
 \sum_{r=0}^{n-1}
 \rho_{m,L,t}^{(\lambda_r)}(x)
 -
 n\rho_{m,L,t}^{(1)}(x).
\end{align}
The overall minus sign follows from the interpolation convention
$V_t=V^{(\Lambda)}+tR^{(\Lambda)}$ and drops out of the absolute-value
criterion in Eq.~\eqref{eq:uv-assumption}.  The ultraviolet problem is
therefore reduced to controlling a concrete replica response, rather than to
approximating the microscopic binary record pointwise by a smooth profile.
The condition in Eq.~\eqref{eq:uv-assumption} is precisely that the omitted
high-frequency potential have vanishing integrated overlap with this response
along a cutoff satisfying
\begin{equation}
 1\ll\Lambda_L\ll L.
\end{equation}
Together with uniform fixed-cutoff asymptotics along the same sequence
$\Lambda_L$, this yields the microscopic scaling relation stated in
Eq.~\eqref{eq:main-result}.

\section{Exact DPP sampling and outcome-resolved XX benchmark}
\label{sec:xx-numerics}

\begin{figure*}[t]
 \centering
 \includegraphics[width=0.96\textwidth]{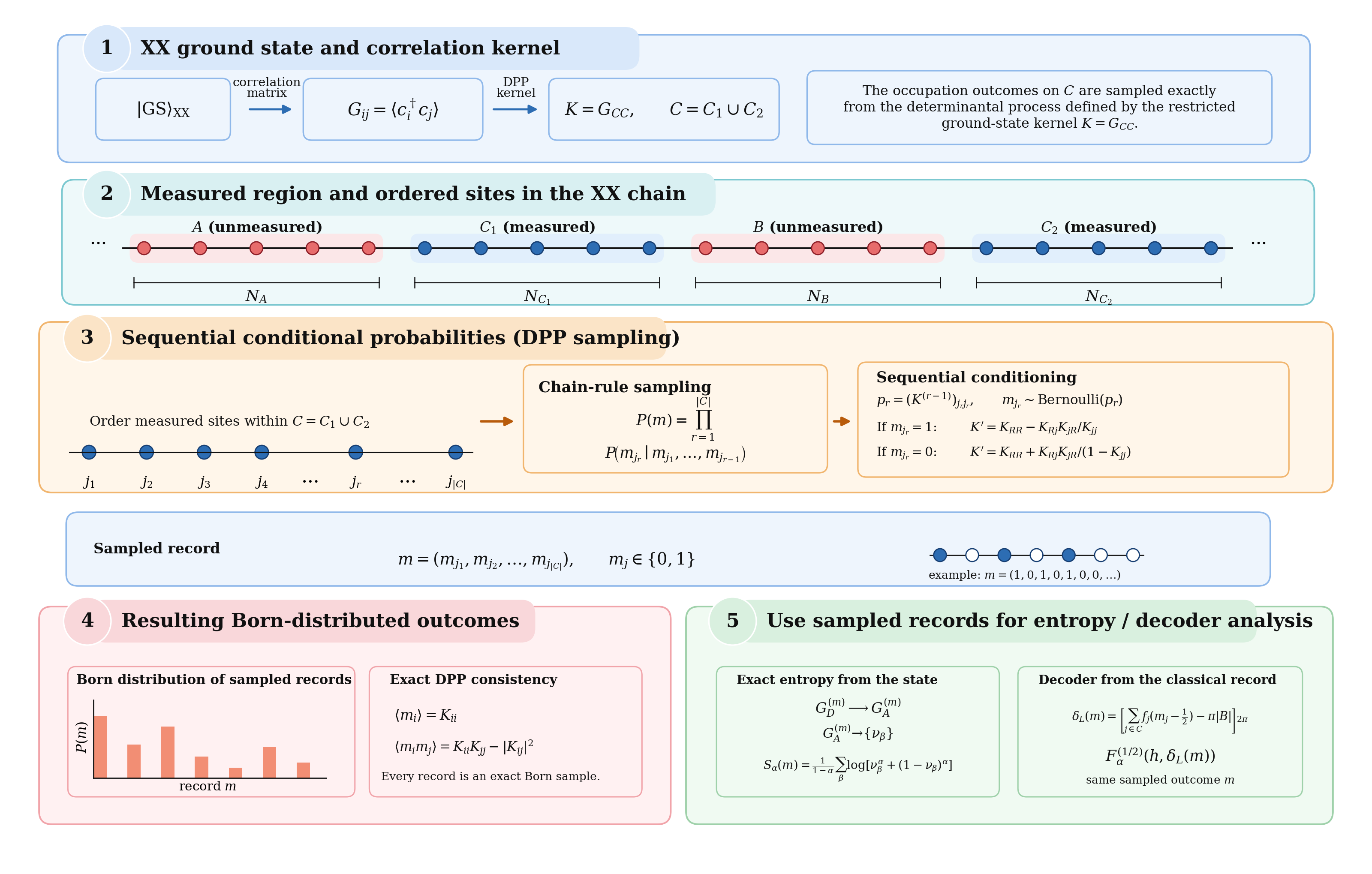}
 \caption{
 Exact Born sampling at the XX point.
 Starting from the ground-state correlation matrix, the measured region
 $C=C_1\cup C_2$ carries the determinantal-point-process (DPP) kernel
 $K=G_{CC}$.  The measured sites are visited sequentially.  At step $r$,
 the occupation of the current site $j_r$ is drawn from the conditional
 Bernoulli probability
 $p_r=(K^{(r-1)})_{j_rj_r}$, after which the kernel on the retained sites is
 updated according to Eqs.~\eqref{eq:dpp-update-one} and
 \eqref{eq:dpp-update-zero}.
 Here $R$ denotes the sites retained after the current draw: the
 as-yet-unsampled measured sites together with the unmeasured subsystem
 $D=A\cup B$.  Thus only one diagonal element of the current kernel is
 needed to draw the next occupation, while the entire retained kernel is
 updated after every outcome.
 Repeating the procedure generates a complete measurement record
 $m=(m_{j_1},\ldots,m_{j_{|C|}})$ with its exact Born probability
 $P(m)=\prod_r P(m_{j_r}\mid m_{j_1},\ldots,m_{j_{r-1}})$.
 Propagating the $D$ block through the same conditional updates yields,
 for that same record, the exact conditioned Gaussian kernel
 $G_D^{(m)}$.  The samples are therefore independent exact Born samples;
 no Markov-chain Monte Carlo is involved.
 }
 \label{fig:xx-dpp-sampling}
\end{figure*}

At the XX point the Born distribution of the measurement record and the
conditional postmeasurement state can both be sampled without Markov-chain
Monte Carlo.  This makes the free-fermion problem an unusually clean numerical
benchmark: every record is drawn with its exact finite-size Born probability,
the entropy of the same record is obtained from its exact conditioned
correlation matrix, and the compact-boson prediction is evaluated independently
from the record coordinate $\delta_L(m)$.  The complete sampling and analysis
workflow is summarized schematically in
Fig.~\ref{fig:xx-dpp-sampling}.  We first spell out the DPP sampling procedure
because it is useful both conceptually and for reproducing the numerics.

% ================================================================
% Exact DPP sampling workflow
% ================================================================

\subsection{Determinantal point process and exact sequential sampling}

Let $G$ be the ground-state Fermi projector and let
$C=C_1\cup C_2$ be the measured region.  The occupation set
\begin{equation}
 O_m=\{j\in C:m_j=1\}
\end{equation}
forms a determinantal point process (DPP) with marginal kernel
\begin{equation}
 K=G_{CC}.
 \label{eq:dpp-kernel}
\end{equation}
Equivalently, for every subset $X\subset C$,
\begin{equation}
 \Pr(X\subseteq O_m)=\det K_X,
 \label{eq:dpp-inclusion}
\end{equation}
where $K_X$ is the restriction of $K$ to $X$.  The first two consequences are
\begin{equation}
 \Pr(m_i=1)=K_{ii},
 \qquad
 \Pr(m_i=m_j=1)=K_{ii}K_{jj}-|K_{ij}|^2.
 \label{eq:dpp-one-two-point}
\end{equation}
Thus the off-diagonal entries of $K$ encode the correlations between measured
occupation outcomes.

The determinant structure turns the joint distribution into an exact
sequential sampler.  Order the measured sites as
$C=(j_1,\ldots,j_M)$, with $M=|C|$.  At step $r$, let
$\mathcal K^{(r-1)}$ denote the current correlation kernel after conditioning
on the already sampled outcomes
$m_{j_1},\ldots,m_{j_{r-1}}$.  The probability that the next site is occupied is
simply its current diagonal element,
\begin{equation}
 p_r
 =\Pr\!\left(
 m_{j_r}=1\mid m_{j_1},\ldots,m_{j_{r-1}}
 \right)
 =\mathcal K^{(r-1)}_{j_rj_r}.
 \label{eq:dpp-next-probability}
\end{equation}
We then draw $m_{j_r}\sim\operatorname{Bernoulli}(p_r)$.

For clarity, define
\begin{equation}
 R_r=D\cup\{j_{r+1},\ldots,j_M\},
 \qquad D=A\cup B,
 \label{eq:dpp-R}
\end{equation}
namely all sites retained after the current measured site $j_r$ has been
conditioned.  The retained kernel is updated by the Schur-complement rules
\begin{align}
 m_{j_r}=1:\quad
 \mathcal K^{(r)}_{R_rR_r}
 &=
 \mathcal K^{(r-1)}_{R_rR_r}
 -
 \frac{
 \mathcal K^{(r-1)}_{R_rj_r}
 \mathcal K^{(r-1)}_{j_rR_r}}
 {\mathcal K^{(r-1)}_{j_rj_r}},
 \label{eq:dpp-update-one}
 \\
 m_{j_r}=0:\quad
 \mathcal K^{(r)}_{R_rR_r}
 &=
 \mathcal K^{(r-1)}_{R_rR_r}
 +
 \frac{
 \mathcal K^{(r-1)}_{R_rj_r}
 \mathcal K^{(r-1)}_{j_rR_r}}
 {1-\mathcal K^{(r-1)}_{j_rj_r}}.
 \label{eq:dpp-update-zero}
\end{align}
Only one diagonal entry is needed to draw the next bit, but after every draw the
\emph{entire retained kernel} is updated.  Consequently
$\mathcal K^{(r-1)}$ already contains the effect of all previous measurement
outcomes, not merely the immediately preceding one.

The origin of Eqs.~\eqref{eq:dpp-update-one} and
\eqref{eq:dpp-update-zero} can already be seen from two sites.  From
Eq.~\eqref{eq:dpp-one-two-point},
\begin{align}
 \Pr(m_j=1\mid m_i=1)
 &=K_{jj}-\frac{|K_{ij}|^2}{K_{ii}},
 \label{eq:dpp-two-site-one}\\
 \Pr(m_j=1\mid m_i=0)
 &=K_{jj}+\frac{|K_{ij}|^2}{1-K_{ii}}.
 \label{eq:dpp-two-site-zero}
\end{align}
These are precisely the diagonal entries of the two conditional kernels above.
Iterating the same identity gives the many-site sampler, and the probability of
the generated record is the ordinary chain-rule product
\begin{equation}
 p_L(m)
 =\prod_{r=1}^{M}
 p_r^{\,m_{j_r}}
 (1-p_r)^{1-m_{j_r}}.
 \label{eq:dpp-record-probability}
\end{equation}
No accept/reject step, equilibration, or autocorrelation time is involved: each
completed bit string is an independent exact Born sample.

For the record distribution alone, retaining the current $C$ block is enough.
In the implementation we simultaneously retain the unmeasured block $D$.  The
same updates in Eqs.~\eqref{eq:dpp-update-one} and
\eqref{eq:dpp-update-zero} then propagate the $C$--$D$ and $D$--$D$ blocks as
well.  After the last measured site has been conditioned, no measured sites
remain and the surviving block is exactly
\begin{equation}
 \mathcal K^{(M)}_{DD}=G_D^{(m)},
 \label{eq:dpp-final-kernel}
\end{equation}
which is the conditioned Gaussian kernel of the postmeasurement state.  This is
numerically equivalent to the closed-form expression
$G_D^{(m)}=G_{DD}-G_{DC}(G_{CC}+M_m-I_C)^{-1}G_{CD}$, but avoids a fresh dense
solve for every sampled record.

\subsection{Two independent routes from the same record}

Each exact Born sample is subsequently processed in two independent ways.  On
the lattice side, we restrict $G_D^{(m)}$ to $A$ and diagonalize
\begin{equation}
 G_A^{(m)}\,u_\beta=\nu_\beta u_\beta,
 \qquad 0\le\nu_\beta\le1.
\end{equation}
For $\alpha\ne1$, the exact finite-size Gaussian entropy is
\begin{equation}
 S_{\alpha,L}(m)
 =\frac{1}{1-\alpha}
 \sum_\beta
 \log\!\left[
 \nu_\beta^\alpha+(1-\nu_\beta)^\alpha
 \right],
 \label{eq:xx-numerical-renyi}
\end{equation}
while the von Neumann entropy is evaluated directly as
\begin{equation}
 S_{1,L}(m)
 =-\sum_\beta
 \left[
 \nu_\beta\log\nu_\beta
 +(1-\nu_\beta)\log(1-\nu_\beta)
 \right].
 \label{eq:xx-numerical-vn}
\end{equation}
Thus the lattice calculation at $\alpha=1$ does not require replica
continuation.

Independently, the same classical bit string is mapped to
\begin{equation}
 m\longmapsto\delta_L(m)
 \longmapsto
 \cF_\alpha^{(1/2)}\!\left(h,\delta_L(m)\right),
 \label{eq:xx-numerical-decoder}
\end{equation}
with $h$ fixed entirely by the subsystem geometry.  For noninteger $\alpha$ and
for $\alpha=1$, the continuum curve in
Eq.~\eqref{eq:xx-numerical-decoder} uses the real-index continuation of the
compact-boson expression described in Sec.~\ref{app:limits}; in particular
$\cF_1=-\partial_\alpha\log\cR_\alpha|_{\alpha=1}$.  No fit parameter is
introduced in comparing Eqs.~\eqref{eq:xx-numerical-renyi}--
\eqref{eq:xx-numerical-vn} with Eq.~\eqref{eq:xx-numerical-decoder}.

% ================================================================
% Outcome-resolved three-alpha benchmark
% ================================================================
\begin{figure}[t]
 \centering
 \includegraphics[width=0.6\columnwidth]{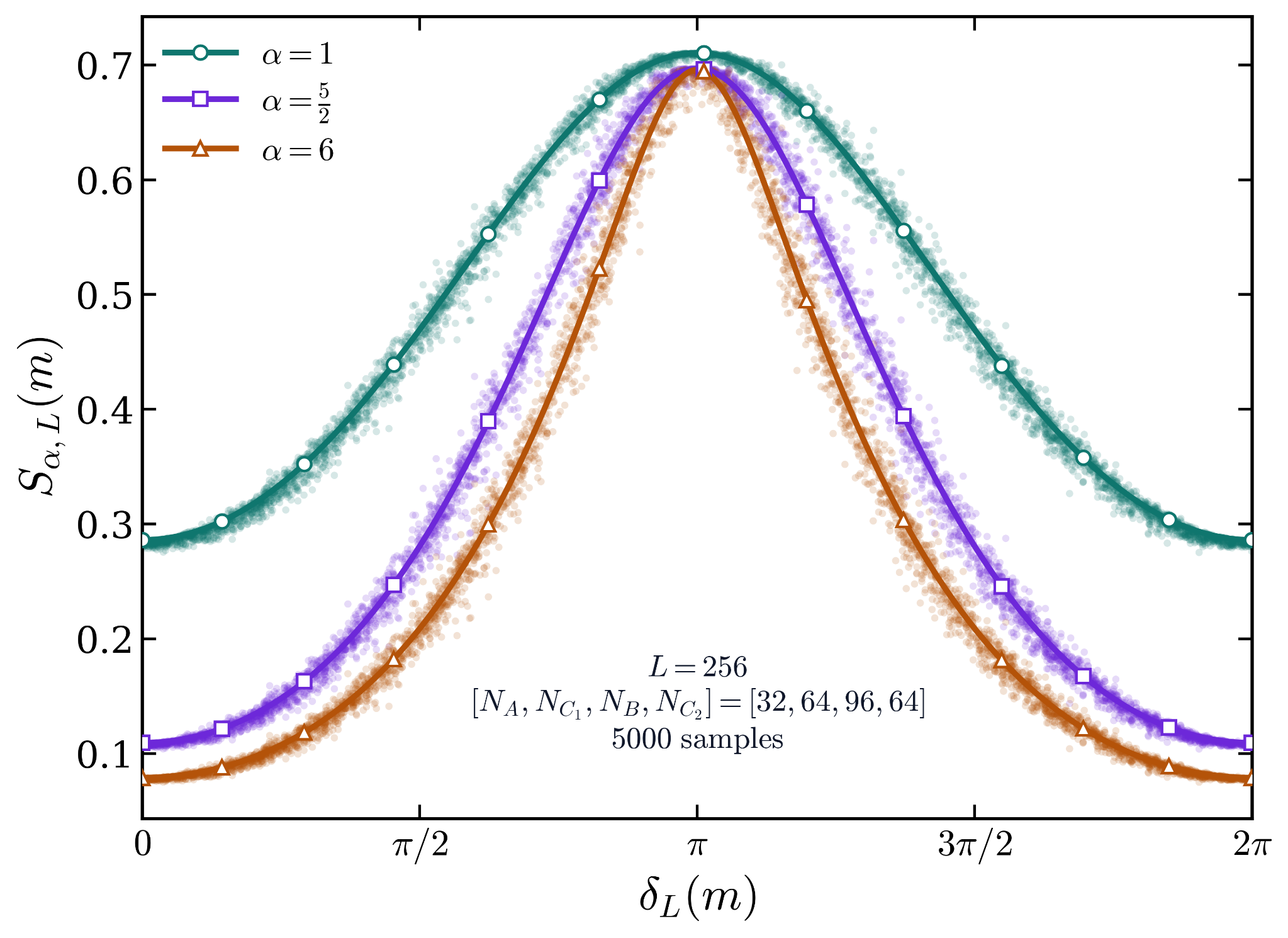}
 \caption{
 Outcome-resolved XX benchmark for three entropy indices.
 Exact conditional entropies $S_{\alpha,L}(m)$ are shown versus the compact
 record coordinate $\delta_L(m)$ at $L=256$ for the strong-asymmetry geometry $[N_A,N_{C_1},N_B,N_{C_2}]=[32,64,96,64]$.
 For each of $\alpha=1$, $5/2$, and $6$, the translucent points are
 $5000$ independent records drawn from the exact DPP Born distribution and
 evaluated using the finite-size conditioned correlation matrix.
 The solid curves are the parameter-free compact-boson predictions
 $\cF_\alpha^{(1/2)}(h,\delta_L)$ for the same geometry.
 The three choices span the von Neumann entropy, a noninteger R\'enyi entropy,
 and a high R\'enyi entropy, and the finite-size outcomes track the
 corresponding one-coordinate prediction across the full compact interval.
 }
 \label{fig:xx-three-alpha}
\end{figure}

\subsection{Numerical protocol and finite-size diagnostics}

The main-text XX figure presents two complementary outcome-resolved tests.
Figure~\ref{fig:xx}(a) fixes $L=256$ and $\alpha=5/2$ and compares two
different geometries, testing whether the same microscopic construction
continues to organize the conditional entropy when both the conformal
coordinate $\delta_L$ and the cylinder height $h$ are changed.
Figure~\ref{fig:xx}(b) instead fixes the strong-asymmetry geometry and compares
the exact entropy directly with
$\cF_\alpha^{(1/2)}(h,\delta_L)$ for $
\alpha=2, \frac{5}{2}, 4.
$
The inset shows the corresponding finite-size decrease of
$\epsilon_{\alpha,L}$ for the same three R\'enyi indices.

The pointwise discrepancy is quantified by
\begin{equation}
 E_{\alpha,L}
 =
 \left\langle
 \left[
 S_{\alpha,L}(m)
 -\cF_\alpha^{(1/2)}
   \!\left(h,\delta_L(m)\right)
 \right]^2
 \right\rangle_{\rm Born},
 \qquad
 \epsilon_{\alpha,L}
 =
 \sqrt{
 \frac{E_{\alpha,L}}
 {\operatorname{Var}_{\rm Born}S_{\alpha,L}}
 } .
 \label{eq:xx-numerical-error}
\end{equation}
The main-text data show that the normalized residual decreases with increasing
system size for the displayed R\'enyi indices.

Figure~\ref{fig:xx-three-alpha} provides a complementary fixed-size test.
At $L=256$ and for the same strong-asymmetry geometry, we compare $
\alpha=1,\frac{5}{2},6.
$
This extends the outcome-resolved comparison from the von Neumann entropy to a
noninteger and a high R\'enyi index.  For $\alpha=1$, the entropy is evaluated
directly from the conditioned one-body eigenvalues $\{\nu_\beta\}$,
\begin{equation}
 S_{1,L}(m)
 =
 -\sum_\beta
 \left[
  \nu_\beta\log\nu_\beta
  +(1-\nu_\beta)\log(1-\nu_\beta)
 \right],
 \label{eq:xx-numerical-vn}
\end{equation}
while the continuum curve is the corresponding
$\alpha\to1$ continuation
$\cF_1^{(1/2)}(h,\delta)$.

The two figures therefore probe different aspects of the same
outcome-resolved relation.  The main-text figure tests geometry dependence and
finite-size convergence, whereas Fig.~\ref{fig:xx-three-alpha} tests the
persistence of the same one-coordinate description across widely separated
entropy indices at fixed size and geometry.  In both cases the comparison is
performed record by record rather than only after Born averaging.

\section{Interacting XXZ numerics and finite-entanglement checks}
\label{app:xxz-numerics}

We detail the interacting calculations supporting the outcome-resolved
comparison in Fig.~\ref{fig:xxz}.  Exact diagonalization (ED) provides
finite-size Born probabilities and conditional entropies without a
tensor-network approximation, while a periodic, $U(1)$-symmetric
matrix-product state (MPS) extends the calculation to larger rings.
For each measurement record, the entropy is evaluated from the
postmeasurement state, independently of the compact coordinate
$\delta_L(m)$ and the parameter-free continuum prediction
$\cF_\alpha^{(g(\Delta))}(h,\delta_L(m))$.  The comparisons
below test the ED and MPS calculations at the same size and then examine
the von Neumann, noninteger R\'enyi, and high-R\'enyi entropies in a larger
MPS calculation.

\subsection{Exact diagonalization and conditional entropies}

We diagonalize the XXZ Hamiltonian, Eq.~\eqref{eq:xxz}, in the
zero-magnetization sector.  For each record $m$ on the measured region
$C=C_1\cup C_2$, projecting the ground-state amplitudes onto $m$
leaves an unnormalized state on $U=A\cup B$.  Arrange its amplitudes
into a matrix $\mathsf A_m$, with row and column indices corresponding
to configurations of $A$ and $B$, respectively.  The record probability
and reduced density matrix are then
\begin{equation}
 p(m)=\operatorname{Tr}(\mathsf A_m\mathsf A_m^\dagger),
 \qquad
 \rho_A(m)=
 \frac{\mathsf A_m\mathsf A_m^\dagger}{p(m)}.
 \label{eq:xxz-ed-conditional}
\end{equation}
Thus the normalized squared singular values of $\mathsf A_m$ are the
Schmidt probabilities $\{p_\beta(m)\}$.  Complete record enumeration
gives exact finite-size Born-weighted observables, subject only to
the numerical accuracy of the diagonalization.  The coordinate
$\delta_L(m)$ is computed separately from Eq.~\eqref{eq:delta};
it is not inferred from the conditional spectrum.

\subsection{Periodic MPS, sequential Born sampling, and Schmidt spectra}

For the MPS calculation we use a two-site $U(1)$-symmetric ansatz with
alternating tensors $A^s$ and $B^s$, whose virtual indices are closed
periodically.  The $U(1)$ charge selection rules keep the state in the
zero-magnetization sector and permit the transfer-matrix and Schmidt
calculations to be organized into charge blocks.  We use the tensors
of the optimized periodic state for the larger-system calculation;
in the same-size ED--MPS benchmark the same tensors are closed on
the shorter ring rather than independently reoptimized there.

Unlike the XX calculation, the interacting sampler does not use a
determinantal point process.  Define the local transfer matrices
\begin{equation}
 T_j=\sum_{s=0,1}T_j^{(s)},\qquad
 T_j^{(s)}=M_j^s\otimes\overline{M_j^s},
 \label{eq:xxz-local-transfer}
\end{equation}
where $M_j=A$ or $B$ according to the site parity.  The periodic
MPS norm is
$\mathcal N=\operatorname{Tr}(T_1T_2\cdots T_L)$.
A prescribed record has Born weight
\begin{equation}
 p_{\rm MPS}(m)=
 \frac{1}{\mathcal N}
 \operatorname{Tr}\!
 \left[
 \prod_{j=1}^{L}
 \begin{cases}
  T_j^{(m_j)},&j\in C,\\
  T_j,&j\in U,
 \end{cases}
 \right],
 \label{eq:xxz-mps-record-weight}
\end{equation}
with the product taken in site order around the ring.  Consequently
we can sample only the measured spins, without first drawing complete
configurations and without Markov-chain Monte Carlo.

Order the measured sites as $j_1,\ldots,j_{|C|}$.  At step $r$,
let $L_{r-1}$ be the transfer product incorporating the preceding
sampled spins and all intervening unmeasured sites, and let
$R_{r+1}$ contract the as-yet-unresolved suffix.  The two candidate
weights are
\begin{equation}
 W_r(s)=
 \operatorname{Tr}\!
 \left[
 L_{r-1}T_{j_r}^{(s)}R_{r+1}
 \right],
 \qquad s=0,1,
 \label{eq:xxz-mps-branch-weights}
\end{equation}
which give the conditional probability
\begin{equation}
 P_{\rm MPS}
 \left(m_{j_r}=1\mid m_{j_1},\ldots,m_{j_{r-1}}\right)
 =
 \frac{W_r(1)}{W_r(0)+W_r(1)}.
 \label{eq:xxz-mps-conditional-prob}
\end{equation}
After drawing $m_{j_r}$, its selected transfer matrix is
incorporated into the prefix.  The resulting record obeys the
chain-rule distribution of Eq.~\eqref{eq:xxz-mps-record-weight}.
The implementation performs these contractions in conserved
transfer-charge-difference blocks; sectors with negligible
full-ring norm contribution are omitted for efficiency, and their
total discarded norm fraction is monitored.  Thus sampling is
independent and Born distributed within the retained MPS
approximation, rather than being exact sampling of the XXZ
ground state.

For each sampled record, fixing the physical indices on $C$
produces the conditional periodic tensor network on $U$.
We form the Gram matrices of the unmeasured blocks and combine
them with the products of the selected measured-site tensors.
The resulting Schmidt problem separates into physical-charge
blocks, whose nonzero eigenvalues give the normalized Schmidt
probabilities $\{p_\beta(m)\}$.  Both ED and MPS then use
\begin{align}
 S_{\alpha,L}(m)
 &=
 \frac{1}{1-\alpha}
 \log\sum_\beta p_\beta(m)^\alpha,
 \qquad \alpha\ne1,
 \notag\\
 S_{1,L}(m)
 &=-\sum_\beta p_\beta(m)\log p_\beta(m).
 \label{eq:xxz-schmidt-entropies}
\end{align}
The noninteger and von Neumann entropies are therefore
calculated directly from the conditional spectrum, not
obtained by continuing a lattice replica formula.

\subsection{ED--MPS comparison and the three-index test}

\begin{figure}[t]
 \centering
 \includegraphics[width=0.6\columnwidth]{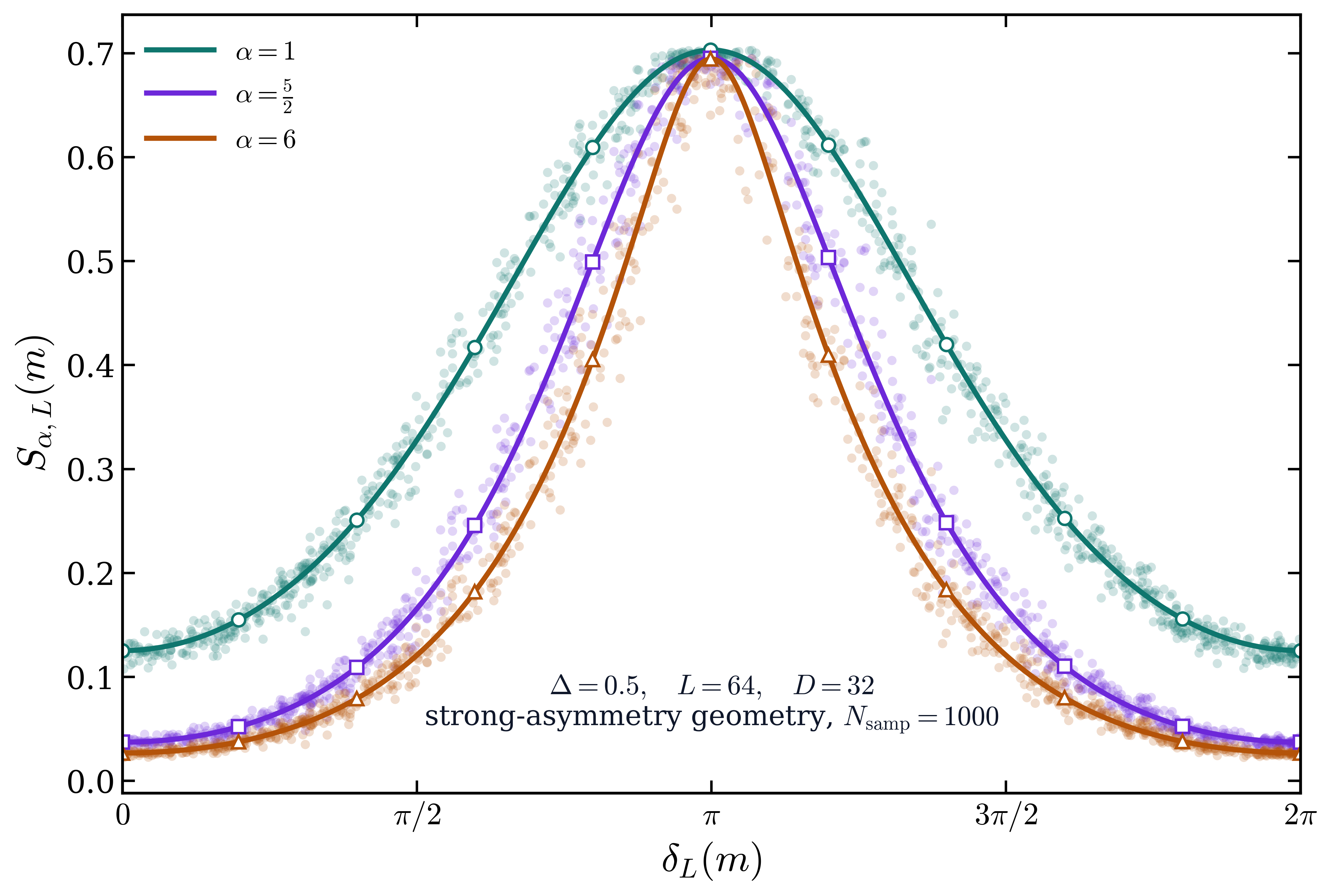}
 \caption{
 Outcome-resolved MPS entropies versus the compact record
 coordinate at $\Delta=1/2$ for
 $\alpha=1$, $5/2$, and $6$.
 The calculation uses $L=64$, periodic bond dimension $D=32$,
 and the strong-asymmetry geometry
 $[N_A,N_{C_1},N_B,N_{C_2}]=[8,16,24,16]$.
 The translucent points represent $1000$ independently sampled
 MPS Born records.  Solid curves and open markers give the
 parameter-free predictions
 $\cF_\alpha^{(2/3)}(h,\delta_L)$ for the same geometry.
 The von Neumann entropy is evaluated directly from the
 conditional Schmidt probabilities.
 }
 \label{fig:xxz-mps-three-alpha}
\end{figure}

Figure~\ref{fig:xxz-mps-three-alpha} provides the complementary
larger-system test.  The three indices probe the von Neumann limit,
a noninteger R\'enyi entropy, and an entropy more strongly
sensitive to the largest Schmidt probabilities.  In each case
the conditional MPS entropy is compared, record by record, with
the same compact-boson function at the known interacting
coupling.  No rescaling or shift of $\delta_L$, additive entropy
offset, or effective coupling is fitted to the points.

\subsection{Finite-entanglement control and scope of the comparison}

The finite MPS bond dimension introduces a correlation length.
For the two-site transfer matrix, whose leading eigenvalues
are $\lambda_1$ and $\lambda_2$, its value in lattice-site
units is
\begin{equation}
 \xi(D)
 =
 -\frac{2}{\log|\lambda_2/\lambda_1|}.
 \label{eq:mps-xi}
\end{equation}
The factor of two converts the two-site unit cell to lattice
sites.  A finite-$D$ calculation at increasing $L$ is therefore
not automatically in the conformal thermodynamic regime;
the ratio $L/\xi(D)$ must also be controlled
\cite{Tagliacozzo2008,Pollmann2009}.  The same-size comparison
in Fig.~\ref{fig:xxz-ed-mps-three-alpha} directly assesses the
finite-$D$ state against ED, while
Fig.~\ref{fig:xxz-mps-three-alpha} tests the conditional
one-coordinate description at a larger size.

The continuum nonzero-mode cancellation holds at general $g$,
but the identification of the interacting lattice coordinate
$\delta_L(m)$ with the compact boundary zero mode is not
established by the free-fermion microscopic argument.  The
interacting data consequently provide numerical tests of this
identification and of its entropy prediction, not a microscopic
XXZ theorem.  An independent check of the coordinate itself is
provided by the compact-boson prediction
\begin{equation}
 P_h^{(g)}(\delta)
 =
 \frac{1}{2\pi}
 \sum_{p\in\bbZ}
 \exp\!\left[
 -\frac{hp^2}{2g}-\ii p\delta
 \right],
 \qquad
 \langle\cos\delta\rangle
 =
 \e^{-h/(2g)},
 \label{eq:general-g-delta-law}
\end{equation}
which involves no conditional-entropy calculation
\cite{KhannaVasseur2026,KhannaVasseurStats2026}.
Together, the ED benchmark, the transfer-matrix sampling
procedure, and the finite-entanglement control delimit the
numerical status of the interacting decoder.

\section{Consequences: Born average, entropy statistics, and geometric limits}
\label{sec:consequences}

Having established the outcome-resolved relation, we now collect several
consequences that follow by combining it with the compact-coordinate
distribution $P_h(\delta)$.  These include the Born-averaged MIE, the limiting
entropy distribution, the short- and long-cylinder limits, and the continuation
to the von Neumann entropy.  No additional microscopic input is required beyond
that entering Eq.~\eqref{eq:main-result}.

\subsection{Born-averaged MIE and the full entropy distribution}

For integer $n\ge2$, the Born-averaged measurement-induced R\'enyi entropy is
\begin{equation}
 \overline S_n(h)=\int_0^{2\pi}\!\dd\delta\,
 P_h(\delta)\,\cF_n(h,\delta).
 \label{eq:mre}
\end{equation}
This recovers the compact-boson Born-averaged structure of
Refs.~\cite{KhannaVasseur2026,KhannaVasseurStats2026}.  The additional content of
the present result is the joint relation: $\delta_L$ is assigned explicitly to
each microscopic outcome and predicts its conditional entropy.

The same statement fixes the complete limiting statistics.  If
$\vartheta\in[0,2\pi)$ is drawn from $P_h$, then
\begin{equation}
 S_{n,L}\ \Longrightarrow\ \mathsf S_{n,h}=\cF_n(h,\vartheta).
 \label{eq:entropy-law}
\end{equation}
Thus the limiting entropy distribution is the pushforward of the compact
coordinate under $\cF_n$.  At a regular entropy value $s$, its density is
\begin{equation}
 p_{n,h}(s)=
 \sum_{\delta:\,\cF_n(h,\delta)=s}
 \frac{P_h(\delta)}{|\partial_\delta\cF_n(h,\delta)|}.
 \label{eq:entropy-density}
\end{equation}
Near a nondegenerate extremum of $\cF_n$, this produces the square-root edge
$p_{n,h}(s)\propto|s-s_\star|^{-1/2}$.

\subsection{Short- and long-cylinder limits}

The geometry enters only through $h$, so the two degeneration limits follow
directly from the compact-boson expression.  Let
$d(\delta)=\min_{w\in\bbZ}|\delta+2\pi w|\in[0,\pi]$.  As $h\to0$, the winding
sum is dominated by its nearest compact sector and
\begin{equation}
 \cF_n^{(g)}(h,\delta)\longrightarrow
 \begin{cases}
 0,&d(\delta)<\pi,\\
 \log2,&d(\delta)=\pi.
 \end{cases}
 \label{eq:short-limit}
\end{equation}
The antipodal value has zero weight in the limiting Born measure, so
$\overline S_n(h)\to0$.

For $h\to\infty$, Poisson resummation and the modular asymptotic of the eta
function give
\begin{equation}
 \cF_n^{(g)}(h,\delta)=
 \frac{h}{12}\left(1+\frac1n\right)
 -\frac12\log(2g)+O(\e^{-c_{n,g}h}),
 \label{eq:long-limit}
\end{equation}
independently of $\delta$ at leading order.  At the XX point $g=1/2$ the
constant vanishes, so the same leading expression holds for the Born average.
The long cylinder therefore retains compact-coordinate fluctuations in the
outcome distribution, while their effect on the entropy is exponentially small.

\subsection{Von Neumann continuation}

The exact lattice factorization in Sec.~\ref{sec:lattice} was derived for
integer $n\ge2$, whereas the conditional correlation matrix defines the lattice
entropy directly for any real index.  The continuum expression
Eq.~\eqref{eq:Rn} has a natural continuation $n\to\alpha>0$.  If the
lattice-to-continuum matching is valid in a neighborhood of $\alpha=1$ and the
thermodynamic limit can be differentiated there, the outcome-resolved von
Neumann prediction is
\begin{equation}
 \cF_1(h,\delta)=
 -\left.\partial_\alpha\log\cR_\alpha(h,\delta)\right|_{\alpha=1}.
 \label{eq:F1-derivative}
\end{equation}
The explicit theta/eta form is given in Appendix~\ref{app:limits}.
Under the same continuation,
$\overline S_1(h)=\int P_h(\delta)\cF_1(h,\delta)\,\dd\delta$, with
$\overline S_1\to0$ for $h\to0$ and
$\overline S_1=h/6+O(\e^{-c_1h})$ for a long XX cylinder.

These ensemble and geometric consequences follow from, but do not replace, the
outcome-resolved statement.  They depend only on the pushforward of the compact
coordinate and therefore do not by themselves imply a unique entropy at fixed
$\delta$.  The stronger infrared structure tested above is
\begin{equation}
 m\longmapsto\delta_L(m)\longmapsto S_{\alpha,L}(m).
 \label{eq:map}
\end{equation}

For the interacting XXZ chain, Fig.~\ref{fig:xxz} directly tests the conditional
map in Eq.~\eqref{eq:map} with the general-$g$ continuum function.  We do not
promote all distributional statements of this section to microscopic XXZ
theorems: unlike at the XX point, the interacting identification of the lattice
snapshot coordinate with the continuum zero mode is presently numerical.  An
independent zero-mode statistic and the tensor-network convergence checks are
reported in Appendix~\ref{app:xxz-numerics}.

\section{Continuum limits and von Neumann formula}
\label{app:limits}

The explicit continuum formulas used in
Sec.~\ref{sec:consequences} are collected here.  We first state the
real-index continuation of the compact-boson replica ratio and its von Neumann
limit, and then summarize the short- and long-cylinder asymptotics.

\subsection{Von Neumann continuation}

For real $\alpha>0$, we use the analytic continuation of
Eq.~\eqref{eq:Rn} obtained by replacing $n$ with $\alpha$,
\begin{equation}
 \cR_\alpha^{(g)}(h,\delta)
 =
 \frac{\eta(q_1)^\alpha}{\eta(q_\alpha)}
 \frac{\Theta_\alpha^{(g)}(h,\delta)}
 {\Theta_1^{(g)}(h,\delta)^\alpha},
 \qquad
 q_\alpha=\e^{-2\pi^2\alpha/h}.
\end{equation}
At $\alpha=1$ the ratio is normalized,
\begin{equation}
 \cR_1^{(g)}(h,\delta)=1,
\end{equation}
and the corresponding von Neumann prediction is
\begin{equation}
 \cF_1^{(g)}(h,\delta)
 =
 -\left.
 \partial_\alpha
 \log \cR_\alpha^{(g)}(h,\delta)
 \right|_{\alpha=1}.
 \label{eq:app-F1-derivative}
\end{equation}

At the XX point, $g=1/2$, let
\begin{equation}
 q=\e^{-2\pi^2/h},
 \qquad
 a=\frac{\delta}{2\pi},
\end{equation}
and define the winding-sector expectation
\begin{equation}
 \langle X(w)\rangle_{h,\delta}
 =
 \frac{
 \displaystyle
 \sum_{w\in\bbZ}
 X(w)\,
 q^{\frac12(w+a)^2}
 }{
 \Theta_1(h,\delta)
 }.
 \label{eq:app-winding-expectation}
\end{equation}
Using
\begin{equation}
 q\partial_q\log\eta(q)
 =
 \frac{E_2(q)}{24},
 \qquad
 E_2(q)
 =
 1-24\sum_{k\ge1}\frac{kq^k}{1-q^k},
\end{equation}
Eq.~\eqref{eq:app-F1-derivative} gives
\begin{equation}
 \cF_1(h,\delta)
 =
 \log\!\frac{\Theta_1(h,\delta)}{\eta(q)}
 +
 \frac{2\pi^2}{h}
 \left[
 \frac12
 \left\langle
 (w+a)^2
 \right\rangle_{h,\delta}
 -
 \frac{E_2(q)}{24}
 \right].
 \label{eq:app-F1}
\end{equation}
This expression is real, smooth, even in $\delta$, and $2\pi$ periodic.
In the lattice calculation, the von Neumann entropy is obtained independently
from the conditioned one-body spectrum; Eq.~\eqref{eq:app-F1} is therefore the
continuum continuation tested against that exact finite-size quantity.

\subsection{Short- and long-cylinder limits}

For $h\to0$, the winding sum is dominated by the sector minimizing
\begin{equation}
 |\delta+2\pi w|.
\end{equation}
For a generic $\delta$ there is a unique minimizer, and therefore
\begin{equation}
 \cF_n^{(g)}(h,\delta)\longrightarrow0.
\end{equation}
At the antipodal point $\delta=\pi$, two winding sectors are degenerate and
\begin{equation}
 \cF_n^{(g)}(h,\pi)\longrightarrow\log2.
\end{equation}
This is the short-cylinder limit quoted in
Eq.~\eqref{eq:short-limit}.

For $h\to\infty$, Poisson resummation gives
\begin{equation}
 \Theta_n^{(g)}(h,\delta)
 =
 \sqrt{\frac{h}{2\pi gn}}
 \sum_{p\in\bbZ}
 \exp\!\left[
 -\frac{hp^2}{2gn}
 \right]
 \e^{\ii p\delta}.
 \label{eq:app-poisson}
\end{equation}
The corresponding modular asymptotic of the eta function is
\begin{equation}
 \eta(q_n)
 =
 \sqrt{\frac{h}{\pi n}}
 \exp\!\left[
 -\frac{h}{12n}
 \right]
 \left[
 1+O\!\left(\e^{-2h/n}\right)
 \right].
 \label{eq:app-eta-long}
\end{equation}
Combining Eqs.~\eqref{eq:app-poisson} and
\eqref{eq:app-eta-long} with Eq.~\eqref{eq:Rn} yields
\begin{equation}
 \cF_n^{(g)}(h,\delta)
 =
 \frac{h}{12}
 \left(
 1+\frac1n
 \right)
 -
 \frac12\log(2g)
 +
 O\!\left(\e^{-c_{n,g}h}\right),
\end{equation}
which is Eq.~\eqref{eq:long-limit}.  The leading term is independent of
$\delta$: in the long-cylinder limit, the compact coordinate continues to
fluctuate between outcomes, but its effect on the conditional entropy is
exponentially suppressed.

\end{document}